\documentclass[prl]{revtex4}%
\pdfoutput=1

\usepackage{epsfig}
\usepackage{graphics}

\begin{document}
\bigskip
\hskip 4in\vbox{\baselineskip12pt \hbox{FERMILAB-PUB-10-036-A-T}  }
\bigskip\bigskip\bigskip

\title{Interferometers as Probes of  Planckian  Quantum Geometry}

\author{Craig J. Hogan}
\affiliation{  University of Chicago and Fermilab}

\begin{abstract} 
A  theory of position of massive bodies is proposed that results in an observable quantum behavior of  geometry at the Planck scale, $t_P$. 
Departures from classical world lines in flat spacetime are  described by Planckian
noncommuting  operators for 
 position in different directions, as defined by interactions with null waves.   The resulting evolution of position wavefunctions   in two dimensions displays a new kind of directionally-coherent quantum noise of transverse  position.  The  amplitude of the effect in physical units is predicted with no parameters, by equating the number of degrees of freedom of  position wavefunctions on a 2D spacelike surface with the  entropy density of a black hole event horizon of the same area. 
    In a region of size $L$, the effect resembles spatially and directionally coherent random transverse shear deformations on timescale $\approx L/c$ with typical amplitude $\approx \sqrt{ct_PL}$.   This quantum-geometrical ``holographic noise'' in position is not describable  as fluctuations of a quantized metric, or as any kind of fluctuation, dispersion or propagation effect in quantum fields.
In a Michelson interferometer the effect appears as noise  that resembles a random Planckian walk of the beamsplitter for durations up to the light crossing time. 
 Signal spectra and correlation functions in interferometers are derived, and predicted to be comparable with  the sensitivities of current and planned  experiments. 
 It is proposed that nearly co-located Michelson interferometers of laboratory scale, cross-correlated at  high frequency, can test the Planckian noise prediction with current technology. 
 \end{abstract}
\maketitle
\section{Introduction}

In all experimentally tested models of  systems that display quantum behavior, spacetime   is described using classical geometry. Worldlines of  particles are quantized paths on a classical spacetime manifold, and  quantum fields  are functions of  classical spacetime  coordinates.  Although this theoretical approach agrees with experiments even at the highest energies,  effects of gravity render it theoretically inconsistent  beyond the Planck scale, $t_P \equiv\sqrt{\hbar G_N/c^5}= 5.39\times 10^{-44}$ seconds. So far, this scale has been out of reach of experiments.
This paper presents theoretical arguments for a new Planckian quantum behavior of  geometry,  and proposes an experimental program to test it.

The features that define classical spacetime--- pointlike events on a continous manifold, with positions described by a continuous mapping of those points onto real numbers--- are not easily reconciled with the quantum nature of matter and energy.       ``Position''  in quantum mechanics  is not  a coordinate of an event, but  a property of an interaction between bodies or particles, represented mathematically by a self-adjoint   operator.   That is, some quantum operators represent the positions of interactions, and  in the classical limit, these observable operators (apparently) behave like event positions related by a classical metric.  The position of an event cannot itself be a quantum observable, since events do not interact.
Thus, even  such a seemingly simple and intuitive concept as position requires a  theory to connect quantum mechanics with spacetime.  In this sense,  no fundamental quantum theory of position is known, and well-tested hybrid approaches,  such as quantum field theory,  become inconsistent at the Planck scale.

 It has been suggested that in  a fully quantum description of the world, classical spacetime itself somehow  emerges as a  limiting behavior of  a quantum system that includes both spacetime and matter. A  description of  nonclassical behavior of position observables in this system can be called a ``quantum geometry''. 
  It has also been suggested that in such a theory (e.g., \cite{Banks:2011av,Banks:2011jt}), the metric itself should not be quantized, since it is itself an emergent classical entity.  That idea is also the starting point here. The metric is treated classically, causal structure is preserved, and light obeys standard physics, but we posit a  noncommutative quantum geometry for position operators and wavefunctions of massive bodies, and evaluate some observable consequences.     It is posited that positions and rest frames in spacetime emerge from quantum physics in a particular way: interactions of null fields with matter define spacetime position in each direction,  position operators in different directions do not commute at the Planck scale, and  time evolution corresponds to an iteration of Planckian  operators. 
 It is shown that an experiment using correlated interferometers can provide  experimental clues about this kind of quantum geometry--- either a detection of effects caused by new Planck-scale quantum degrees of freedom, or a Planckian upper bound that   constrains theory.  

\subsection{Motivation}
 It has long been established that the quantum mechanics of physically realizable measurement systems, such as clocks, limits the precision with which classical observables, such as the interval between events described by the classical metric, can be defined
\cite{wigner,salecker,peres, braginsky,aharonov,zurek}.  For some measurements, a  complete account of the quantum system should include trajectories over an  extended region of  space-time.   Well established measurement theory  does not however account for the quantum mechanics of spacetime itself, and how it might be entangled with real-world observables.

It is well known that when gravity is included, spacetime dynamics itself  poses limitations on  any physically realizable clock. At the Planck scale, even the separation of quantum and space-time concepts becomes inconsistent: matter confined to a box smaller than the Planck scale in all three dimensions lies within the Schwarzschild radius for its quantum-mechanically expected mass, causing a singularity in the spacetime; on the other hand, a black hole smaller than the Planck length does not even have enough mass to make up a single quantum at its Schwarzschild frequency.  Because of this inconsistency, some kind of new physics must enter that effectively imposes  a  maximum frequency at the Planck scale.  A universal Planck frequency  bound imposes a new kind of uncertainty on the definition of spacetime position that applies to any physically realizable measurement apparatus\cite{Padmanabhan:1987au}.     Although it is acknowledged that fundamentally new, quantum spacetime physics occurs at the Planck scale, the  physical character of  Planckian position uncertainty is not known and has been inaccessible to experimental tests. 

Some features of quantum geometry  have been understood precisely and consistently  from a blend of relativity, thermodynamics and field theory. The Bekenstein-Hawking entropy of a black hole, which  maps into degrees of freedom of emitted particles during evaporation, is given by one-quarter of the area of the event horizon in Planck units.   It has been proposed that this result generalizes to  a 2D Planckian holographic encoding of quantum degrees of freedom in any spacetime.   According to this ``Holographic Principle'', spacetime quantum degrees of freedom  can  be covariantly described in terms of a boundary theory, with a Planckian information  density on surfaces defined by the boundaries of causal diamonds\cite{'tHooft:1993gx,Susskind:1994vu,Jacobson:1995ab,Bousso:2002ju,Padmanabhan:2009vy,Verlinde:2010hp}.  A holographic theory must depart substantially from a straightforward extrapolation of conventional quantum field theory, both in the number of degrees of freedom and in the notion of locality.  However, there is no agreement on the character of those degrees of freedom---  their physical interpretation, phenomenological consequences, or experimental tests. 

There are other rigorous mathematical approaches to nonclassical spacetime physics,  such as  noncommutative geometry\cite{Seiberg:1999vs,connes,connesbook}. Quantum conditions imposed on spacetime coordinates change the algebra of functions of space and time, including quantum fields and position wavefunctions. Instead of quantizing the metric or fields directly,  position operators are quantized. For some classes of commutators, and some physical interpretations in terms of quantum fields, these geometries have been constrained by experiments\cite{mattingly}. Once again however, at present there is no experimental evidence for departures from classical geometry that could guide the physical interpretation of the theory.

In this situation it makes sense to conduct experiments that explore outcomes outside the predictive scope of currently well-tested physical theory, whose results will help to guide the creation of a quantum  theory of geometry. 
This paper presents a simple wave theory  of  position of massive bodies in flat spacetime to define a macroscopic limit of quantum geometry.  The wave theory incorporates  features of emergence, holography, and noncommutative geometry, and predicts specific new observable  behaviors in the macroscopic limit.  In particular the theory is used to predict   a new kind of uncertainty in relative transverse position that leads to noise in  interferometers, corresponding to a Planck amplitude spectral density of fluctuations in transverse position.     The predictions  can be   tested with  current technology.

\subsection{Description}

In  many widely considered theories, new Planckian physics does not create  any detectable effect on  laboratory scale positions of bodies. For example, in a straightforward application of field theory to spacetime modes, quantum fluctuations on very small scales   average to unobservable amplitude in  measurements of position in much larger systems.  However,  this approach may not be the correct low-energy, large-scale effective theory to describe  new Planckian physics. 
The effective theory  described here posits  quantum conditions that preserve classical coherence and Lorentz invariance in each direction, but  departs from the standard commutative behavior of positions in different directions.  In this framework, Planckian effects become detectable.

The main hypotheses  here are that interactions of null fields with matter define spacetime position in each direction; that position operators in different directions do not commute at the Planck scale; and that time evolution corresponds to an iteration of Planckian  operators.  As a result,  transverse uncertainty in spacetime position measurements accumulates over macroscopic distances, instead of averaging rapidly to zero.  Although the classical limit is well defined and is not changed, the new behavior changes the  approach to the classical limit and produces a larger position  uncertainty on macroscopic scales than field theory predicts on its own. This new kind of spacetime position indeterminacy has precisely calculable statistical properties, and leads  to a new kind of noise in nonlocal comparative measurements  of transverse relative positions on macroscopic scales, for example in interferometers. The rough overall magnitude in a laboratory scale experiment is an attometer-scale jitter on timescales of tenths of microseconds--- very small, but likely detectable.
  The main new feature required to detect it is that the interferometer signals should be recorded and correlated at a rate comparable with the inverse light travel time for the apparatus. This requires an unusual experimental setup, but no fundamental breakthrough in technology.

The  new macroscopic behavior can be roughly characterized in several equivalent ways. Transverse positions of trajectories separated by distance $L$ appear to fluctuate by amount $\Delta x \approx \sqrt{ct_PL}$ on a timescale $\approx L/c$.  Regions of size $L$ appear to undergo coherent random shear deformations in rest frame velocity on the same timescale, with typical amplitude $\Delta v \approx c \sqrt{ct_P/L}$. On longer timescales $\tau> L/c$, the relative angular positions $\theta$ of matter trajectories fluctuate coherently by about $\Delta \theta/\theta\approx \sqrt{t_P/\tau}$. These phenomenological descriptions refer to nonlocal optical measurements of position in a macroscopic system, extended in  time and two dimensions of space. However, they derive from Planck-scale physics and have the character of quantum noise.  Precise statistical predictions  of its behavior are derived below, specifically for  cross-correlated signals of nearly co-located  Michelson interferometers.

\subsection{Relation to previous theories and experiments}

 This macroscopic behavior  has a distinctive  phenomenology, qualitatively different from several other proposed Planckian or Lorentz-invariance-violating effects that have been analyzed using tools of effective field theory\cite{mattingly}.    The new uncertainty and noise are associated purely  with mean macroscopic spacetime position and velocity, independent of any parameters of effective field theory, or indeed any parameters apart from the Planck scale.  
    For example, this effect adds no  dispersion to  particle propagation, and is invisible to such tests proposed for cold-atom interferometers\cite{AmelinoCamelia:2009zz}.  
 It would also have no dispersive effect on cosmic photon propagation: null particles of all energies in any one direction are predicted to propagate in the usual way at exactly the same velocity, in agreement with current cosmic limits, from Fermi/GLAST satellite observations of gamma-ray bursts, on the difference of propagation speeds at different photon energies\cite{fermi2009}.  Similarly, no new effect is predicted for  energy dependence of polarization position angle, consistent with  INTEGRAL/IBIS satellite bounds\cite{Laurent:2011he}.   

  The   noise in interferometers predicted here also behaves differently from  Planckian noise previously predicted from quantum-gravitational or metric fluctuations, quantization of very small scale spatial field modes, or spacetime foam\cite{Ellis:1983jz,AmelinoCamelia:1999gg,AmelinoCamelia:2001dy,Schiller:2004tf,Smolin:2006pa,Ng:2000fq,Ng:2004xr,Lloyd:2005dm}. Indeed, many of these ideas are either now ruled out by data,  or remain far out of reach of experiments.   By contrast,   the effect discussed here would heretofore have escaped detection, yet is measurable with current technology. To avoid confusion with the earlier ideas, it is sometimes useful to adopt the term ``holographic noise'' 
  to refer to the  effect described here, which depends on a new Planckian uncertainty in position of matter in a fixed classical metric. 
The most conspicuous  difference in physical behavior is that  no holographic noise appears in measurements of position in a single direction, unlike noise from  a fluctuating metric. Measurements of holographic noise display coherent quantum correlations associated with  entanglement of position states in overlapping regions of spacetime that cannot be described as a fluctuating metric, because they derive from a definition of position based on noncommuting  operators.  As discussed below, these features lead to  effects in interferometers that differ from metric fluctuations, and depend on the details of the experiment design. 
    
 The new physics proposed here violates Lorentz invariance, but in a specific way that has not been previously tested.  There is no  causality violation, although there is a  new kind of quantum-mechanical entanglement of systems that share a common spacetime volume, even if there is no other physical connection between them. There is also no preferred frame or direction, except that which is set by the measurement apparatus. Indeed, the effective theory here defines a particular way that a classical rest frame could emerge from a quantum theory.  The effect can only be detected in an experiment that coherently compares transverse positions over an extended spacetime volume to extremely high precision, and with high time resolution or bandwidth. One reason that the effect of the fluctuations is strongly suppressed in most laboratory tests is that over time, average positions approach their usual classical values; as noted above, the apparent fractional distortion from classical geometry in a system of size $L$ is predicted to be of order $\sqrt{t_P/\tau}$  for   measurements averaged over time $\tau> L/c$.

 Of course, interferometers also have a well-understood standard quantum noise limit  that follows directly from the Heisenberg uncertainty principle, or equivalently, from quantization of electromagnetic field modes\cite{caves1980}. A measurement of arm-length difference $X$ for duration $\tau$, with a minimum mirror mass $m$, is uncertain by
 \begin{equation}\label{SQL}
\Delta X^2= 2\hbar\tau/m,
\end{equation}
independent of the wavelength of the light.   This limit is based on standard physics: classical spacetime, with massive bodies  described by quantum mechanics and photons described by quantum fields.  (The intensity of the light does not enter explicitly in this formulation, but is accounted for in this bound: a larger mirror mass reacts less to a fluctuating photon momentum and therefore allows a lower  photon shot noise).

 The conjecture here is roughly that new  physics of quantum geometry imposes a new fundamental limit, corresponding to a Planck mass for $m$ in Eq.(\ref{SQL}),  that applies to the precision of measurements of transverse  position. That is,   the quantum physics of position in emergent spacetime somehow imposes a Planckian frequency limit on the spacetime wavefunctions of massive bodies when measured by comparing interactions with null fields in different directions. This new collective behavior represents a departure from standard physics, where a massive body in general has a position state described by a  wavepacket that includes super-Planckian frequencies. 
 
 Such an effect would not have been previously detected. The standard quantum noise of  interferometer signals  can be viewed\cite{caves1980} as interference of zero-point fluctuations of electromagnetic vacuum  state modes (entering from the dark port) with incoming light. Planckian noise in the signal would signify an entanglement of those modes with a new Planckian indeterminacy of the apparatus+spacetime configuration state. The effective wave theory below suggests a  quantitative model of how this might work.

 Some properties of  holographic noise were previously estimated\cite{Hogan:2007pk,Hogan:2008zw,Hogan:2008ir,Hogan:2009mm}, using a different theory also based on position wavefunctions and wavepackets. States were represented as modulations of a fundamental carrier with Planck frequency, evolved with a paraxial wave equation, and the position uncertainty appeared as a diffraction-like effect of Planck carrier waves of the position wavefunction. The new effective wave theory  derived here results in a different effective wave equation, based on  noncommutative deformations on a 2D spacelike surface in a laboratory rest frame where positions are measured.   The new theory  addresses  position uncertainty on 2D spacelike surfaces at rest, as opposed to the 2D spacelike null wavefronts in the earlier description.  In this view, the uncertainty arises from  complementarity of transverse position and rest-frame velocity. This rest-frame perspective appears likely to be more useful for calculations of response in general interferometer configurations.    The two descriptions encode a similar holographic information content and display holographic uncertainty of a similar  amplitude.   In both descriptions, positions in spacetime are encoded with a Planck bandwidth limit, $\approx 10^{44}$ bits per second, and the noise can be viewed as the corresponding Shannon sampling noise of position---  a consequence of a fundamental bandwidth limit on spacetime  position observables. 

\subsection{Relation to quantum gravity theories}
The rest of this paper entirely neglects effects of gravity or of spacetime curvature. Nevertheless, it is important to comment on possible connections with  quantum gravity and emergent spacetime.

Jacobson\cite{Jacobson:1995ab}, Verlinde\cite{Verlinde:2010hp} and others have advanced arguments for a  theory of gravity and inertia based on general thermodynamic and holographic principles. Spacetime is emergent, and positions are encoded on two dimensional  surfaces; gravity is identified as an entropic force, and acceleration is identified with the temperature of a quantum  system that ultimately emerges as matter in spacetime.  The theory here does not address gravity or acceleration directly, since it deals with flat spacetime, corresponding to zero temperature. However, it does present an effective theory that describes  the character of the holographic degrees of freedom of position of matter in emergent holographic spacetime, and a specific kind of coarse graining that relates positions with macroscopic separation.  The fluctuations predicted here could then be a direct experimental  signature of the degrees of freedom whose statistical behavior  gives rise to classical gravity.

Banks\cite{Banks:2011av,Banks:2011jt} has proposed a quantum theory of emergent holographic spacetime based on the following construction: ``A time-like trajectory gives rise to a nested sequence of causal diamonds, corresponding to larger and larger intervals along the trajectory. The holographic principle and causality postulates say that the quantum mechanical counterpart of this sequence is a sequence of Hilbert spaces, each nested in the next as a tensor factor.''   Banks  proposes a matrix theory to describe this quantum system, which is general enough to include gravity and particle states. In that holographic space-time, as here, ``the metric of space-time is encoded in  the relations between various quantum Hilbert spaces and is not itself a fluctuating quantum variable.'' 

The effective theory here addresses only part of this physics, corresponding roughly to the behavior of averages over traces of the matrices\cite{Hogan:2008ir}. It describes the quantum kinematics of averages over many particles, the mean position of massive bodies in emergent holographic flat spacetime, and includes none of the rich dynamics of particles or gravity. It does however capture a similar ``angular delocalization'' of particle states\cite{Banks:2011jt}.
The ``nested system of causal diamonds'' is also closely related to the wave theory developed here and helps to explain its predicted correlations, in particular, its spatial coherence.   In our case, the Hilbert space is defined by spatial position wavefunctions in two dimensions, with a Planck frequency limit.  

\begin{figure}[b]
 \epsfysize=5in 
\epsfbox{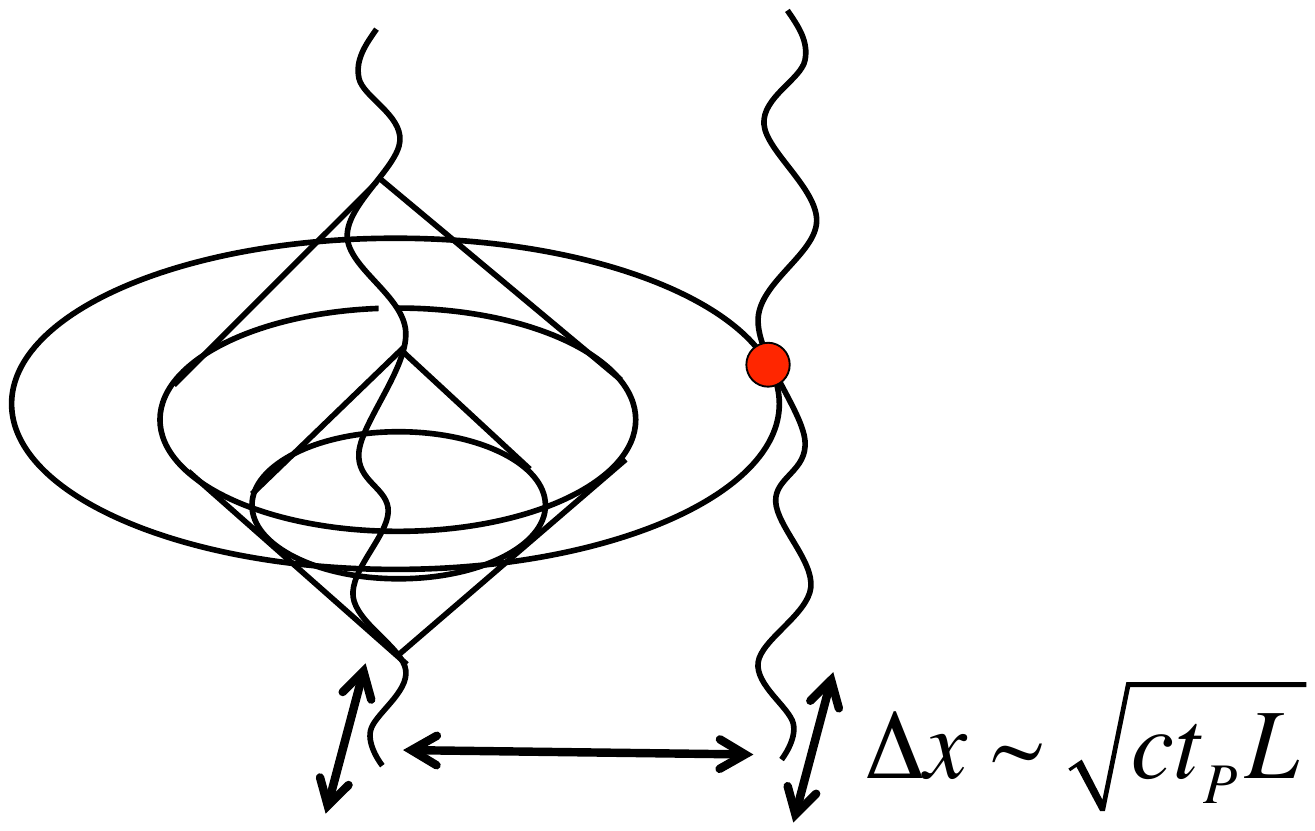} 
\caption{ \label{nesteddiamonds}
Spacetime diagram of the nested-causal-diamond construction associated with collective quantum states of matter position and rest frame in flat classical spacetime.  One spatial dimension is suppressed. The two vertical wavy lines  represent timelike trajectories or world lines of bodies.  A causal diamond is the intersection of the future light cone of one point on a world line with the past light cone of some future point. Two nested causal diamonds associated with the left trajectory are shown. 
It is conjectured that relative  positions of  two world lines in spacetime are encoded by Planck-limited wavefunctions on 2D boundaries of the diamonds tangent to the world lines; one of these 2D surfaces is shown as a spacelike circle intersecting the right world line at the position of the solid dot.  Their rest-frame separation $L$ determines the amplitude of coherent random transverse fluctuations in measured position of amplitude $\Delta x \approx \sqrt{ct_PL}$ on a timescale $\approx  L/c$. }
\end{figure}

\begin{figure}[b]
 \epsfysize=5in 
\epsfbox{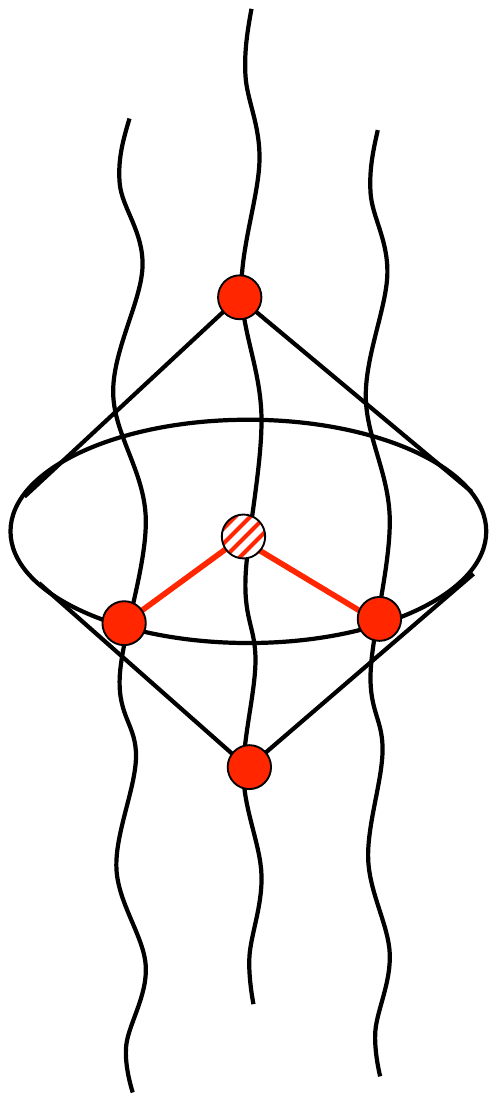} 
\caption{ \label{interferometerdiamond}
Spacetime diagram showing a causal diamond associated with a Michelson interferometer. The spatial dimension out of the interferometer plane is suppressed. The central worldline represents the beamsplitter, the other two represent end mirrors. The two arms are shown on one spacelike surface, a particular time in the lab frame. The measured signal compares light reflected from the two end mirrors, in different directions, as a function of time. The interactions with the beamsplitter in the two directions occur at two times separated by $2L/c$, where $L$ denotes the arm length. The solid dots represent reflection events that contribute to the  signal at the time represented by the uppermost dot. The measurement compares positions nonlocally, and in different directions. The wavepacket description of  uncertainty in an interferometer  refers to wavefunctions in the positions of a single beamsplitter-mirror trajectory at two different times, relative to end mirrors in two different directions.}
\end{figure}

To illustrate  this correspondence, suppose an observer on a timelike trajectory sends out a Planckian series of sonar-like pulses.  The trajectory is defined quantum-mechanically by the set of causal diamonds traced by these  null waves and their received counterparts. The new uncertainty described here appears in the mutual quantum relationship of the position of two different observers, on different trajectories. In particular, their relative transverse positions, as measured by Planckian waves (or Planckian pulses) display a diffractive uncertainty that is much larger than the Planck scale (see Figure \ref{nesteddiamonds}).  The Michelson interferometer system performs a similar measurement that compares a single beamsplitter trajectory at two times in two different directions (see Figure \ref{interferometerdiamond}). In this case the position is continuously measured, so the uncertainty shows up as noise in a signal stream.  The coherence of holographic noise in two nearby interferometers can be understood because of the overlap of  their causal diamonds entangles their quantum geometrical states (illustrated in Figure \ref{topview});  measurements of position  collapse the corresponding wavefunctions into the same state.

Noncommutative geometries\cite{Seiberg:1999vs,connes,connesbook}
 and some of their observational consequences\cite{mattingly} have been extensively discussed in the literature.   The   new features added in the discussion below are a particular physical interpretation of position operators, a particular choice of    commutator, and a particular hypothesis for the time evolution of the system. The physics of nonclassical geometry as interpreted here differs significantly from the more familiar context of field theory deformed by a Moyal algebra. Quantum conditions here are imposed on the 2D position of massive bodies as measured by interactions with null fields in their rest frame, which leads to different physical results from usual quantization of field configuration states.   Moyal  deformations are  applied here not to 3D fields, but   to two dimensional position wavefunctions. Repeated deformations are assumed to generate  time evolution. In this way the  evolution of a quantum geometry and its macroscopic effects can be described using an effective wave theory.  Of course, we do not know what effective equations describe real quantum geometry, but the point here is to present a precisely formulated effective theory that can be quantitatively tested with realizable experiments.

\section{Position in quantum geometry}

Quantum position operators should  automatically obey the causal structure defined by the classical metric.  One way to guarantee  this causality is to posit a quantum definition of a position measurement  in terms of null fields, such as light. For example, a position in a particular direction can be defined by a directional eigenstate of a null field, a plane wave completely delocalized in the transverse directions.

Consider an idealized world consisting of matter and radiation in  an unperturbed, 3+1-dimensional spacetime.
We wish to establish an operational definition of position for matter. For definiteness, consider a reflecting surface. It forms a spacelike boundary condition   for an electromagnetic field. Its position is defined by its effect on the field, which is how the position is measured:  the field solution depends on the position of the surface. This system  is classical: neither the surface, nor the field, nor the metric are quantized.  Since the position measurement can include a large area that averages over many atoms, we can take the surface to be perfectly smooth.    The  field in vacuum obeys the standard classical relativistic wave equation, and propagates in a flat classical metric. The vacuum solutions of the field can be decomposed in the usual way into plane wave modes.    These modes are not quantized, so we are not here considering quantization or photon noise in position measurement. 

Position in each direction is measured by a normally-reflected mode traveling in that direction.
The position of a body is defined by measurements based on configurations of  reflected radiation.  The quantum geometry we seek to study is introduced by imposing quantum conditions connecting the position operators in different directions. 

This definition of position manifestly agrees with standard space-time in the classical limit. By construction, measurements in any single direction rigorously respect causality, and exactly agree with classical behavior. However, transverse positions no longer share a single classical space. Comparisons of position in different directions have a quantum relationship that departs from classical behavior by a small amount, depending on the value of the commutator. As shown below, this particular way of implementing a quantum geometry of transverse position results in a surprisingly large departure from classical behavior on macroscopic scales.  

\subsection{Position Operators}

In the rest frame of any body, choose any direction in space.  This direction defines a plane, which we identify as an observer's choice of holographic projection.  In this plane, let  $x_i(t)$ denote the classical position of the body in two dimensional Cartesian coordinates ($i=1,2$).  The correspondence between classical  and  quantum positions is posited to obey  the following quantum commutation relation:
\begin{equation}\label{noncommute}
[\hat x_i, \hat x_j]= i (Cct_P)^2 \epsilon_{ij},
\end{equation}
where $\epsilon_{ij}$ is the unit $2\times 2$ antisymmetric matrix, $\epsilon_{ij}=-\epsilon_{ji}=1$.
The scale is set by the Planck time $t_P$, with a coefficient $C$ of the order of unity that can in principle be normalized from gravitational entropy arguments.

This choice of rectilinear basis vectors is convenient for the  calculations that follow. However,  from linear projection of the position operators and basic trigonometry, one can show that the same physical prescription can be stated in a way that is manifestly independent of the choice of coordinates.  The position operator for a direction inclined by angle $\theta'$ relative to axis 1  is 
\begin{equation}
\hat x(\theta')= \hat x_1 \cos(\theta') +\hat x_2 \sin(\theta').
\end{equation}
For any two directions, the commutator is then
\begin{equation}
[\hat x(\theta'),\hat x(\theta'')]=\{\cos(\theta')\sin(\theta'')-\sin(\theta')\cos(\theta'')\}[\hat x_1, \hat x_2]=\sin(\theta'-\theta'')[\hat x_1, \hat x_2]= i  \sin(\theta'-\theta'') (Cct_P)^2.
\end{equation}
Therefore, the quantum condition (\ref{noncommute}) can be stated independently of coordinates: {\em In the rest frame of a body, the commutator of position operators in any two directions is proportional to the sine of the angle between them, with a Planck scale coefficient.}   This construction makes it clear that the operators defining the new noncommuting geometry do not actually define any preferred frame or direction in 3+1D space, except  for those necessarily determined by a particular measurement, as is usual in quantum mechanics.  That is, the new physics itself does not single out a frame or direction, so it preserves Lorentz invariance in this sense.

Thus, any measured component of a body's position is a quantum operator that does not commute with measurement of orthogonal position components. Because angle is itself frame dependent, the commutator of position operators in two fixed directions does depend on the rest frame of the massive body whose position is being measured, but this is to be expected, since the new physics is connected with  measurement of a rest frame, together with spacetime, as an emergent  structure. In the emerged 3D space, Eq. (\ref{noncommute}) can be written $ [\hat x_i, \hat x_j]= i (Cct_P)^2 \epsilon_{ij3}$, where $\epsilon_{ijk}$ is the completely antisymmetric tensor, and the axis 3 is defined as the normal to the  plane defined by the two measurement directions in the rest frame of the body. This formulation does violate parity symmetry; the sign of the commutator changes on reflection in the holographic plane.
 
It is important that the new Planckian behavior is associated with directions in which positions are measured. 
A plane wave exactly aligned with a planar reflecting surface reflects in an exactly classical way;  no new physics is detectable.  Thus, a one-dimensional optical cavity that compares phases of waves reflecting between parallel surfaces detects no new nonclassical effect, to first order.  On the other hand, the phase of a reflected  plane wave with orientation inclined to the surface depends on position components in different (incoming and outgoing) directions, and these do not commute. The state of the (otherwise classical) radiation field is affected by the (quantum, and Planck bandwidth limited) state of the boundary condition. 

Indeed, nothing about photon propagation in vacuum is changed by adding the commutator, Eq.(\ref{noncommute}). The electromagnetic field  still behaves as in  a perfect classical spacetime with no new Planckian physics. The metric is not perturbed; the new effect is thus not the same as gravitational waves, or any quantization of a field mode.  However, this classical spacetime on its own is not directly accessible to an actual position measurement. That requires interaction with matter at some position, and also a  particular choice of frame and measurement direction.
The position of the boundary condition with matter is where the new Planckian quantum behavior enters: 
it applies to the position of matter in the spacetime, as opposed to the unaltered metric.  The boundary condition affects the radiation field in the usual way, so the configuration of the radiation field depends on the matter position state (and depends on the quantum position operator) even though its equation of motion in vacuum and the metric itself are not changed.

Even though this formulation is based on classical spacetime,  radiation and matter, we have added a new quantum condition on the spacetime positions of matter,  which affects the radiation via interactions.  The system can be placed by interaction into different states. 
We can thus speak of a measurement in a particular direction placing a whole system of matter and radiation into an eigenstate of that direction.  A measurement of a definite, measurable macroscopic configuration state of the field ``collapses the wavefunction'' in the usual way. In this situation, the relative transverse position is not fixed classically until the radiation is detected, which may occur a macroscopic interval away. This holographic nonlocality does not violate causality, but it does correspond to a new kind of uncertainty in position that is shared coherently by otherwise unconnected bodies.

As noted previously,   the usual one dimensional wave equation is obeyed in each direction, and vacuum field modes propagate in the usual way. 
However, quantum operators that measure spacetime intervals, say by comparing ticks of a physical clock with the phase of a wave travelling between events,  have an orientation in space.   If the operators in different directions do not commute, a fundamental limit follows on the accuracy of  position measurements compared in different spatial directions over macroscopic intervals.   A  new source of  noise appears in devices that compare phases of null fields that propagate in different directions, at high frequencies (comparable to the inverse light travel time), across a macroscopic system extending in two spacelike dimensions.  The noise resembles an accumulation of transverse Planck scale position errors over a light crossing time. The new behavior appears as a new kind of transverse jitter or displacement from a classical position.

\subsection{Effect on Interferometers}

The optical elements and detectors of an interferometer create particular  boundary conditions for the radiation field that make this effect detectable, if it exists.
 In a simple Michelson interferometer, light propagates along two orthogonal directions, say, $x_1$ and $x_2$, along arms of length $L$. A single incoming wavefront is split into two noncommuting directions for a time $2L$.  Light enters the apparatus prepared with a particular phase and orientation; the final signal depends on the position of the beamsplitter in two directions,  at two different times separated by $2L$. When recombined the relative phases of the wavefronts have wandered apart from each other by $X \approx \sqrt{2CLct_P}$, just as if the beamsplitter had moved by this amount. The apparent motion is due to Planckian uncertainty in the position and velocity of the beamsplitter. 
 
 In a simple Michelson interferometer, the signal at the dark port represents a measurement of the arm length difference, measured by reflections off the beamsplitter that occur at two different times, in the two directions, separated by an interval $2L/c$. In terms of the position operators introduced above, if we ignore any motion of the end mirrors, the interferometer continuously measures a quantity represented by the operator
 \begin{equation}\label{measurementoperator}
\hat X(t)= \hat x_2(t)-\hat x_1 (t-2L/c).
\end{equation}
An ongoing measurement thus combines two noncommuting operators at two macroscopically separated times.
Moreover, note that Eq. (\ref{measurementoperator}) holds only in the rest frame of the beamsplitter. In a measurement that is distributed in time, an uncertainty in transverse position can be reinterpreted as an uncertainty in transverse velocity and therefore, in the definition of rest frame. The phase of the detected light depends on the relative velocity at the time of the two reflections. The reflection events are shown in Figure (\ref{interferometerdiamond}):  the two arms determine the directions 1 and 2, the end mirrors fix the directions in which positions are measured, and the signal records the difference of the beamsplitter  position at the two times.

A consistent description of the uncertainty including the position and transverse-velocity uncertainty is given below, in terms of wavepackets.
The result is that for  time 
differences  $\tau$ much smaller than  $2L/c$,   there is noise  in the phase comparison of the light from the two arms,    equivalent to a variance in beamsplitter position $\sigma_X^2(\tau)= C c^2  t_P\tau$ at time lag $\tau$.
For larger time differences  $\tau>2L/c$, the  phase does not continue to drift apart, since the wavefronts from the two directions are not prepared in  the same way as plane wavefronts from infinity.  They are not actually independent, but constrained by the finite apparatus size.  The beamsplitter has  a definite position at every time that fixes the relative $x_1$ and $x_2$ phases at a time interval $\tau= 2L/c$. Phase differences at intervals  $\tau>2L/c$ thus represent independent samplings of a distribution about the classical position.    The distribution  has a variance $\sigma^2= 2C L  ct_P $, with a mean that approaches the classical expectation value of arm length difference. It is important to note that the  noise is in nonlocal comparison of relative spacetime position and rest frame averaged over many particles, rather than the position of individual quantum particles. 
 

The construction using directional position operators  suggests that the effect is spatially and directionally coherent.
It seems quite strange that the positions of bodies in a given rest frame and a given direction share the same holographic ``displacement'', even if there is no  physical connection between them. This feature can be traced to the idea that the  commutator is introduced in terms of directional operators that  are independent of transverse position.  In the classical situation, with zero commutator, positional coherence is of course taken for granted; everything has zero holographic displacement. That classical coherence is preserved for nearby paths sharing the same direction.  The holographic displacements depart from the classical behavior by adding  a new transverse jitter that only becomes apparent between paths with a significant transverse separation.   If two parallel paths are much longer than the transverse separation between them, they will measure almost the same total transverse displacement when compared with a  much  longer  transverse path (as in Figure \ref{topview} (b)). The  mean square displacement difference grows linearly with transverse separation. This is a consequence of the displacement occurring transversely relative to light sheets, rather than in three dimensions relative to a fixed laboratory rest frame.  

\begin{figure}[b]
 \epsfysize=5in 
\epsfbox{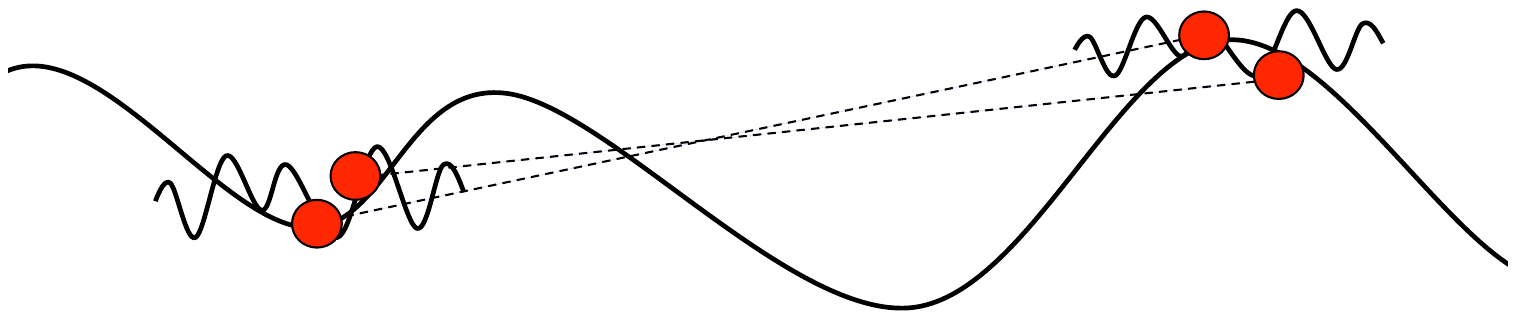} 
\caption{ \label{coherence}
Sketch showing the scaling of Planckian uncertainty and noise. The relative amplitude of position fluctuations is shown from typical wave components on two scales. Dots represent positions of bodies. Vertical displacement shows the (greatly exaggerated) difference from classical position; horizontal scale is time, or distance in the rest frame.     The typical excursion grows with duration like $\sqrt{\tau}$, therefore,  the total displacement is dominated by the longest scale measured, even though the angular fluctuations, here shown by the slopes of the waves, are largest on small scales (ultimately reaching unity on the holographic Planck scale).   The longest wavelength measured corresponds  to the scale of a measured causal diamond, which determines the overall excursion of measured amplitude fluctuations. Spatially overlapping diamonds collapse into the same modes on this scale, so nearby bodies share correlated, directionally coherent motion on scales much larger than their transverse separation, as indicated by dashed lines. These spatially and directionally coherent fluctuations from a classical geometry are shared by collections of particles and bodies in the same region of spacetime. Over timescales long compared with the size of a region, the fluctuations average away to become negligible. }
\end{figure}

The coherence  can also be traced to the scaling property that the amplitude of the holographic  jitter grows with scale (see Figure   \ref{coherence}).  Once again, the effect is different from microscopic quantum fluctuations, which average out in a macroscopic system. Indeed, this averaging is the key to reducing ordinary quantum shot noise enough to allow  macroscopic phase measurements in an interferometer with such precision. The coherence is needed for holographic jitter to be detectable at all; entire macroscopic optical elements of the interferometers ``move'' almost coherently, so the effect  is not reduced by averaging over a macroscopic patch of a mirror surface.
 It is also the reason that holographic noise has escaped detection up to now; it has a smaller amplitude on small scales, yet in a fixed spatial region, averages to zero over long measurement times.

\section{Wavefunctions  in  quantum geometry}

 The new behavior can be described using quantum wave mechanics.  A trajectory in a classical spacetime may resemble  a ray approximation to  a deeper theory based on waves.  We seek a theory  for the waves that captures the same holographic uncertainty just described using operators. This quantum wave theory of position is still only a theory of departures from   classical behavior; it is far from being a fundamental theory of emergent spacetime. However, it suffices to make new experimental predictions.

In classical geometry, quantum wavefunctions of position in each direction are independent.  A noncommutative geometry changes their functional relationship so that they are no longer independent.  The effective theory here is based on propagating the noncommutativity of the the geometry to the algebra of wavefunctions.  A joint wavefunction  that describes  position in two directions at the same time is a product of the wavefunctions in each  direction. This product inherits a new quantum algebra from the quantum conditions of position operators, that describes the  difference from the classical product.  We posit a time evolution for  the joint wavefunction derived from  this difference equation. This conversion amounts to a hypothesis about how the emergent quantum mechanics of spacetime works, in particular, the emergence of position, time, and rest frame in holographic theory. Along with the definition of position in terms of directional null operators, it is the main conceptual assumption being tested by the proposed experimental program.

\subsection{Wave Equation Derived from Moyal Deformation}

We start with the functional deformation caused by noncommutative geometry, described by a Moyal algebra\cite{Seiberg:1999vs,connes}.  Geometry described by $[\hat x_i,\hat x_j]= i \theta_{ij}$  leads to a deformation in the algebra of functions $f$ and $g$, to leading order,
\begin{equation}\label{moyal}
(f * g) - fg = (i/2) \theta_{ij} \partial_i f\partial_j g.
\end{equation}
Such a deformation applied to fields in three dimensions leads to effects at the scale set by $\theta_{ij}$.  In the case of  a Planckian commutator in 3D, such a small effect is not detectable. In particular, if the functions $f$ and $g$ are quantum fields, the geometric uncertainty is confined to the scale of the commutator.  This behavior is similar to the effect of a Planckian UV cutoff in field modes.

The observable effect proposed here 
results from a different, holographic physical interpretation of Moyal deformation.  The new Planckian physics gives rise to  a new, effective wave equation that describes the the position of matter  in two spacelike directions.
Instead of applying the Moyal deformation to the metric, or to quantum fields, as usually done, we apply it at a more primitive level, to describe the noncommutativity of position operators and deformation of wavefunctions.
The relevant functions to use in Eq. (\ref{moyal}) are now not quantum fields, but position wavefunctions in two spatial dimensions.

Consider as above any two orthogonal directions 1 and 2 in the rest frame of the body being measured.  Suppose that the position of the body in each direction is a quantum-mechanical amplitude represented by a wavefunction, $\psi_1(x_1),\psi_2(x_2)$.  We again define positions physically in terms of interactions with directional null modes, so the undeformed wavefunctions in each direction have transverse coherence associated with plane waves,
\begin{equation}\label{orthogonal}
\partial_2\psi_1(x_1) = \partial_1\psi_2(x_2) = 0,
\end{equation}
 to leading order in $ct_P/L$, where $L$ is the characteristic size of the system.
 
We again adopt a Planckian commutator of positions in the Cartesian $x_1,x_2$ plane  given by Eq.(\ref{noncommute}),
\begin{equation}\label{noncommute2}
[\hat x_1, \hat x_2]= i 2\ell_P^2.
\end{equation}
This leads to a Moyal deformation 
\begin{equation}\label{moyal2}
(\psi_1 * \psi_2) - \psi_1  \psi_2 = i \ell_P^2 \partial_1 \psi_1 \partial_2 \psi_2,
\end{equation}
where $\ell_P$ is of the order of $ct_P$.
This can be interpreted as the change in  quantum-mechanical amplitude for the positions $x_1,x_2$  from what they would have been in a nondeformed (classical, commutative) geometry.  In this quantization,  the normalization of the wavefunction is held fixed, and not quantized. Hence there is no ``second quantization'' in the theory, and no  zero point vacuum oscillations. The new effect is a  quantum mechanical uncertainty, not a vacuum fluctuation noise.

As in the 3D case, the 2D positions in  Eq. (\ref{moyal2}) deform from their classical values only by a distance of the order of $\ell_P$. 
Suppose however that this   deformation corresponds to just one Planckian time interval,  a single ``clock tick'' in the   frame defined by the 2D spacelike sheet defined by the directions 1 and 2. The idea is that time evolution is a series of Planckian time displacements: time evolution corresponds to repeated deformations of the form (\ref{moyal2}). To arrive at the effective theory, we approximate the difference equation (\ref{moyal2})  as a continuous differential equation for times much larger than Planck.

These ideas motivate the following evolution equation for the joint, 2+1-D  position wavefunction over times large compared with $t_P$:
\begin{equation}\label{wave}
 \partial_t (\psi_1(x_1,t)\psi_2(x_2,t))= i c \ell'_P \partial_1 \psi_1 \partial_2 \psi_2,
\end{equation}
where $\ell'_P\equiv \ell_P^2\omega_P/c$ is a  spatial scale  closely related to observable noise, and $\omega_P$ denotes a Planckian frequency for the evolution, corresponding to the inverse of the time between  steps as one goes from a discrete to a continuous description. The numerical values of $\ell'_P$ and $\omega_P$ are discussed below.

Equation (\ref{wave}) can be viewed as an effective wave description of holographic modes of emergent spacetime position relative to a particular frame, in the spacelike plane defined by the chosen directions $x_1$ and $x_2$. Note that as usual in quantum mechanics, time itself is not  measured;   positions are measured only in the $x_1,x_2$ plane. Like the Schr\"odinger equation,  Eq. (\ref{wave}) respects linear unitary time evolution required of quantum mechanics.  Unlike the Schr\"odinger equation, it includes a product of derivatives in two spacelike directions that are normally independent.  

Clearly  Eq. (\ref{wave}) has not been derived from fundamental theory in a rigorous way.
Here, we simply posit this equation, in the spirit of the Bohr atom model, as an effective wave theory to describe the new physics,  without pretention to be a fundamental theory. It describes a new, wavelike behavior of position and velocity of matter in spacetime, caused by new Planckian physics,  with some new, if very small effects on larger scales. The behavior reflects  a particular implementation of quantum geometry.  The Planckian time sampling leading to Eq. (\ref{wave}) imposes a  bandwidth limit on the evolution of spatial wavefunctions from one 2D spacelike surface to the next--- in effect,  imposing a Planckian fidelity limit on relative positions in different directions, at different times. It is this effective equation that can be tested in experiments.

The solutions  of Eq.(\ref{wave}) can be written as a combination of modes in the two directions:
\begin{equation}\label{1modes}
\psi_1(x_1,t)=\sum_k A_1(\omega,k_1) \exp[i(\omega t- k_1x_1)],
\end{equation}
\begin{equation}\label{2modes}
\psi_2(x_2,t)=\sum_k A_2(\omega, k_2) \exp[i(\omega t-k_2x_2)],
\end{equation}
with a dispersion relation that relates the two sets of coefficients,
\begin{equation}\label{dispersion}
2\omega=-c \ell'_P k_1k_2.
\end{equation}
The new noncommutative physics  appears in the form of the two-dimensional character of the modes described by this dispersion relation. In the joint 2D wavefunction, modes in the two directions are not independent. 

Equation (\ref{dispersion}) creates a wide gap between the  frequency and wavenumber scales when wavelengths are much larger than $ \ell'_P$.  To describe a state with a macroscopic extension in time of the order of $\tau$, the $A_{1,2}(\omega,k_{1,2})$'s in the  sums (Eqs. \ref{1modes},\ref{2modes}) must extend to low frequencies, of the order of $\omega\approx \tau^{-1}<< t_P^{-1}$. The dispersion relation (\ref{dispersion}) then shows that typical states have  spatial wavefunctions with significant power in transverse spatial modes on scales much larger than the Planck length. That is, the joint wavefunction of position in the two directions includes nonzero $A$'s where $(k_1k_2)^{-1} \approx c\tau \ell'_P>>(ct_P)^2.$  The  time evolution of the emergent position operator thus leads to effects on  a much larger scale than   Planck. The eigenstates have the character of waves with one macroscopic longitudinal dimension (associated in this case with the unmeasured time and space dimensions) and two much smaller, but still not negligible, transverse dimensions. For long durations $>>t_P$, the width is negligible compared to the duration and typical position-state wavepacket trajectories approximate classical  worldlines.

   This description shows the departure from the decomposition standard in field theory, into quantized 3+1-D plane-wave modes. A plane-wave eigenmode in a particular direction, say $k_1$, is now not an exact eigenmode. True eigenmodes include both dimensions, so the plane wave states are entangled with each other. The wave solutions of equation (\ref{wave}) in different directions are not independent of each other, as they are in field theory.

 \subsection{Wavepacket Description of Planckian Position Uncertainty in Interferometers}

The new uncertainty can be understood physically in terms of the width of quantum-mechanical wavepackets. Normally, with a dispersive evolution equation wavepackets  spread with time.  On the other hand, Equation (\ref{wave}) is linear when each direction is considered on its own. There is no dispersive effect observable in a 1D measurement.  But once we choose a   direction for the basis states of the wave expansion (that is, with coefficients $A_{1,2}$ both expressed in terms of $\omega,k_1$ or $\omega,k_2$, with the wavenumber in the other  direction, $k_2$ or $k_1$, fixed by the dispersion relation), the transverse direction wavepacket has an uncertain transverse velocity.
For each $k_1$ mode, the dispersion relation (\ref{dispersion})  associates it with a velocity in direction $x_2$:
\begin{equation}
v_2=d\omega/d k_2 = -c \ell'_P k_1/2.
\end{equation}
 An eigenmode of wavenumber in direction 1 maps onto a transverse velocity in direction 2, so a measurement of position in   direction $x_1$ (say) creates uncertainty in $k_1$, and hence in transverse velocity $v_2$.
The same statement applies with $1,2$ reversed. A wavepacket with a spread of $k_1$'s necessarily has a spread of $v_2$'s (and vice versa).
This effect represents the essential element of the new physics of the uncertainty: a state with a  position wavepacket in one direction has a conjugate uncertainty in wavenumber, and therefore also in transverse wavenumber and  velocity, and hence a phase uncertainty that accumulates with transverse propagation. 

The new uncertainty can be illustrated using a Michelson interferometer as a concrete example.
A Michelson interferometer measurement combines two terms (Eq. \ref{measurementoperator}) that  correspond to position-space wavepackets at two times, in two directions (see Fig. \ref{interferometerdiamond}). Denote the wavefunctions at the two reflections by $\psi_1(x_1,t) $ and  $\psi_2(x_2,t+2L/c)$, and their standard deviations  by $\Delta x_1(t) $ and  $\Delta x_2(t+2L/c)$. 
In wavenumber space, the wavepacket of the first reflection  has a standard deviation $\Delta k_1= 1/\Delta x_1$.  
  The reflected light interacts with matter that has an effective transverse velocity $v_2= c \ell'_P k_1$, which is uncertain by  
\begin{equation}
\Delta v_2= c\ell'_P \Delta k_1/2= c\ell'_P/2\Delta x_1.
\end{equation}
  After a time $2L/c$ the velocity leads to a phase shift  of the reflected light, with a standard deviation  in length units
  \begin{equation}
  \Delta x_2 = 2\Delta v_2L/c=   L \ell'_P/\Delta x_1.
  \end{equation}   
  The phase-difference observable $X=x_1-x_2$ has a wavefunction whose variance  is the sum of two terms that depend oppositely on $\Delta x_1$:  
  \begin{equation}
\Delta X^2 = 
\Delta x_1^2+\Delta x_2^2=
 \Delta x_1^2 +   ( L \ell'_P/\Delta x_1)^2.
\end{equation}
The minimum uncertainty for the measurement of $X$ occurs when  the two terms are equal, $ \Delta x_1^2 = ( L \ell'_P/\Delta x_1)^2= L\ell'_P.$   The probability distribution for the difference measurement has a standard deviation
\begin{equation}\label{wavepacketuncertainty}
 \Delta X= \sqrt{2L \ell'_P},
\end{equation}
 which is $>>\ell'_P$. As discussed below, over shorter time intervals $\tau<L/c$, the position-difference observable displays fluctuations or noise with excursions of amplitude $\Delta X\approx \sqrt{c\tau \ell'_P}$. 
 
 The spread in the frequency-space wavepacket corresponds to a new measurement uncertainty in the definition of a rest frame: a measurement of position in one direction leads to velocity uncertainty in the transverse direction. In addition to position uncertainty of a measurement, there is a new transverse Planckian velocity uncertainty and a corresponding uncertainty in phase that grows with propagation distance.

 \subsection{Uncertainty, measurement noise, nonlocality, and correlations}

 As usual, the quantum theory describes the evolution of a wavefunction.  It makes only statistical predictions about outcomes of experiments.  In a real experiment, the uncertainty and indeterminacy described by the quantum wavefunction manifests as randomness or noise.
 
 The   behavior on short timescales  resembles clock error: position difference wanders by about a Planck time per Planck time.  
However, the wave description shows that the effect should not be viewed simply as a  random walk, but is due to the complementary uncertainty of  transverse rest-frame velocity and position of the 2D wavefunctions.   In particular, it is also not right to think of the effect as random walk in the orientation of  light rays. Indeed,  light propagation has been regarded as classical all along; it is the position of rest frames that is uncertain.  The uncertainties in angular position and rest-frame velocity become smaller  on larger scales  (as in Fig. \ref{coherence}), so directions become more classical, and larger systems more precisely approximate matter inhabiting a classical three dimensional space.  However, the transverse position-difference uncertainty increases with scale, up to the size of the causal diamond defined by an apparatus. Thus, the longest waves dominate the amplitude of the fluctuations.  These occur coherently over a light-crossing time.
 
A measurement in an interferometer determines a definite value  at any given time for the position-difference observable. 
  In the causal diamond picture,  randomness continually enters the system from the world outside the causal diamond, causing decoherence of the overall quantum wavefunction. Measurements sharing the same causal diamond, in the same orientation, must ``collapse the wavefunction'' to the same spacetime state at each time, because they measure the same incoming random modes at the same time and place. Although this  interpretation of quantum mechanical measurement uncertainty is standard\cite{zurek}, the application to positions defining a quantum geometry is new.  There is no  violation of the causal structure of classical spacetime.  However, correlations arise, in the quantum departure from classical behavior, between different systems that have no physical connection apart from sharing the same spacetime volume.
   
Detection of the effect depends on a measurement apparatus with macroscopic spacelike extent in  two directions.
For experiments, the nonlocal character of the states provides a powerful diagnostic technique using cross correlation. Two nearly co-located and co-aligned interferometers that share an overlapping volume of spacetime, but otherwise have no physical connection, experience common mode  fluctuations, since the wavefunctions of the spacetime volumes they measure must collapse into the same state--- the same coefficients $A$ for modes on the scale of the apparatus. Most of the displacement for an apparatus of size $L$ is from modes with wavelengths of order $L$, and to the extent their measured diamonds overlap, their states are strongly correlated over time intervals of order $L/c$. If they are offset or misaligned from each other, the cross correlation is reduced, and if they probe nonoverlapping spacetime volumes, the correlation vanishes altogether.  The effects of these correlations on observable signals are estimated quantitatively below.

The full quantum description of the interferometer will include the quantum geometry discussed here, as well as the degrees of freedom of the particle content. In the standard theory of interferometer noise\cite{caves1980}, the normal modes of the photon field are modified by mirror reflections. The photon states (number eigenstates) have a delocalized character in space, extending across the entire apparatus. When the quantum geometry is taken into account, these modes are entangled with the position states of the matter in the mirrors. The derivation of modes above adopts an approximation, that the beamsplitter position-difference observable at each time is measured using position defined by photon states prepared at infinity. In reality the states are shaped by the apparatus, in particular, the  measurement  depends on the relative position of the beamsplitter and end mirror trajectories. Although a full quantum theory of the entangled system is not worked out here,   signal correlations in a real experiment are estimated below, based on constraints imposed by causality.

\subsection{Relation to Black Hole Entropy}

It is  instructive  to compare the spacetime degrees of freedom encoded on spacelike surfaces of the effective wave theory  with the entropy of a black hole event horizon.  This is the most direct way to set an absolute normalization for the effective theory, and thereby for experimental predictions.

The spacetime modes here are described in flat spacetime. The treatment   breaks down for systems (or modes) whose size approaches the radius of spacetime curvature. For an experiment on the Earth's surface, that is about a light hour ($\approx c (G_N\rho)^{-1/2}$ where $\rho$ denotes the mean density of the Earth), so curvature can be safely neglected in description of any laboratory apparatus. Curvature of a null wavefront corresponds to a gravitational focusing of normal rays, and it is this gravitational lensing that links the thermodynamic description of spacetime to the classical Einstein equations\cite{Jacobson:1995ab}.

In the case of a black hole, the curvature radius corresponds to the Schwarzschild radius.  Modes on this scale exhibit  Hawking radiation, which converts the spacetime degrees of freedom into particle degrees of freedom whose excitation is detectable far from the hole.  Curvature of the event horizon connects long wavelength modes to modes outside the horizon that appear to a distant observer (or in flat space, an accelerating one) as thermally populated.   This thermal conversion process cannot be described using  the flat-space theory described by equation (\ref{wave}).
However, we conjecture that the  number of  degrees of freedom is the same on any 2D spacelike surface, whether in a laboratory or a black hole.

  Consider  modes on a rectangular 2D spacelike surface, with sizes $L_1$ and $L_2$ in the two dimensions at rest. These lengths set the maximum wavelength and minimum wavenumber of modes on the surface, $k_{1min}=2\pi/L_1, k_{2min}=2\pi/L_2$, and modes of each have integer multiples of these minimum values.
Suppose also that there is  a Planckian maximum frequency, $\omega_{max}$,  in the effective theory. 
The number of degrees of freedom is identified with number of independent modes--- the number of different $k_1,k_2$ pairs consistent with the dispersion relation (Eq. \ref{dispersion}), with frequency up to  $\omega_{max}$.

We  choose to count by $k_1$.  At each $k_1$,  values of $k_2$ are integer multiples of $k_{2min}$. The frequencies have values  given by Eq. (\ref{dispersion}), up to  a maximum given by the maximum frequency, $\omega_{max}$. At $k_{1min}$ the maximum is
\begin{equation}\label{k2max}
k_{2max}= 2\omega_{max}/c\ell'_Pk_{1min}.
\end{equation}
The total number of modes ${\cal N}$ is given by summing over discrete values of $k_1$ from $k_{1min}$ to $k_{1max}$, a sum that has $k_{1max}/k_{1min}$ terms:
\begin{equation}\label{surfacelog}
{\cal N}= {k_{2max}\over k_{2min}} \left(1+ {1\over 2} + {1\over 3}+ \dots + {k_{1min}\over k_{1max}}\right)
\approx {2 \omega_{max} A\over (2\pi)^2 c\ell'_P} \log(k_{1max}/k_{1min}),
\end{equation}
where $A=L_1L_2= (2\pi)^2/k_{1min} k_{2min}$ is the area of the surface. 

The log factor in this counting may well be unphysical; it arises from modes which are super-Planckian in one of the spacelike dimensions.
It goes away if we insist that $k_{1}$  and $k_{2}$ are both  subject to a Planckian upper bound, which seems reasonable for an emergent theory of geometry. This idea explicitly introduces the notion of a maximum frequency of spatial wavefunctions, as mentioned in the introduction in the context of the standard quantum limit for interferometers, Eq. \ref{SQL}.
If we set $k_{1max}= k_{2max} =\omega_{max}/c$, the number of modes is
\begin{equation}\label{surfacemax}
{\cal N}= \left({k_{1max}k_{2max}\over k_{1min} k_{2min}}\right)
= { A \omega_{max}^2 \over (2\pi)^2 }
= { A   \over (2\pi)^2 }{2\omega_{max}\over c\ell'_P},
\end{equation}
where the last equality is imposed by the dispersion relation, Eq. (\ref{dispersion}). This fixes the maximum frequencies to the scale of the commutator, so there is only one independent scale for the effective theory. That in turn can be normalized by reference to black holes.

Since the effective theory describes ``pure spacetime'' quantum degrees of freedom, it is natural to to identify  the  number of degrees of freedom (Eq. \ref{surfacemax}) with  the  entropy of a black hole event horizon of the same area, 
\begin{equation}\label{blackhole}
S={A\over 4 (ct_P)^2}, 
\end{equation}
where $t_P \equiv\sqrt{\hbar G_N/c^5}$ is the conventional definition of Planck time.
 By setting ${\cal N}= S$ we arrive at a normalization for the effective theory, $\omega_{max}=\pi/t_P$, or
\begin{equation}\label{normalization}
 \ell'_P= 2 ct_P/\pi.
\end{equation}
This formula fixes the  observable noise amplitude   in standard physical units, as in Eq. (\ref{wavepacketuncertainty}).

This estimate is precise, but not necessarily accurate. In the absence of a more complete microscopic theory that connects the wave theory to curved spacetime, we do not know that  ${\cal N}=S$ is an exact relation; the argument lacks  precise control over the correspondence of black hole entropy to position degrees of freedom in the wave theory.   
 A fundamental theory that clarifies these relationships should eliminate the arbitrary character of the assumptions, and can be tested directly and precisely. 
 The concrete estimate here serves as a  suitable target for experimental design, but it should be emphasized that in the real world the absolute normalization might be different by a factor of the order of unity. Other predictions, such as the shapes of the frequency spectrum and cross correlation function, do not depend on this overall amplitude normalization.

\subsection{Relation to Paraxial Wave Theory}
A different equation was previously suggested as the basis of a candidate effective theory\cite{Hogan:2009mm}, partly based on a connection with the kinematics of Matrix theory\cite{Hogan:2008ir}. For a Michelson interferometer with a classical observable quantity  $X= x_1-x_2$,  the wavefunction was posited to obey a 1+1D  paraxial wave equation,
\begin{equation}\label{schro}
 \partial_t \psi(X, t)= - i c^2t_P \partial_X ^2 \psi(X, t),
\end{equation}
which has  wave solutions
\begin{equation}\label{1Dmodes}
\psi(X,t)=\sum_k A_k\exp[i(\omega t- kX)],
\end{equation}
and dispersion relation
\begin{equation}\label{schrodispersion}
\omega=  c^2 t_P k^2.
\end{equation}
Except for the coefficient,  Eq.  (\ref{schro}) resembles the nonrelativistic Schr\"odinger wave equation. Quantum uncertainty based on this equation is described in analogy with wave optics: the wave solutions have a diffractive transverse beamwidth.  For interferometers, there are periodic solutions  for the wavefunction in analogy with optical cavities, where the position uncertainty corresponds to the beam width, and the apparatus size corresponds to the cavity length. These solutions are useful to illustrate the  character of the noise  in a finite apparatus.

The two equations (\ref{wave}) and  (\ref{schro})  refer  to spatial wavefunctions on different kinds of 2D  spacelike hypersurfaces. Eq. (\ref{schro}) describes motion referred to  null wavefronts (in this case, those defined by laser cavity modes), while Eq. (\ref{wave})  describes motion on a 2D surface at rest, in this case the plane of the interferometer. We have not yet investigated the detailed connection between these different views of the effect. Eq. (\ref{wave}) describes  a  conjugate relationship between two transverse directions not present in the Eq. (\ref{schro}): it can ``squeeze uncertainty'' into one direction or another, it is manifestly linear and nondispersive in each direction, and it is motivated here by connection to a time series of  Moyal deformations. Both equations represent a similar information bound, corresponding to the holographic number of degrees of freedom, and display similar macroscopic uncertainty.

\section{Statistical Properties of Holographic Noise}
 
The above properties suffice to estimate the statistical properties of the noise in an interferometer. We express the detected phase  as the apparent arm-length difference $X(t)$, in length units.  We first estimate the time-domain autocorrelation function  for  a single interferometer, defined as
\begin{equation}\label{autocorrelation}
\Xi(\tau)\equiv \lim_{T\rightarrow\infty} (2T)^{-1}\int_{-T}^{T} dt X(t) X(t+\tau)\equiv \langle X(t)X(t+\tau)\rangle.
\end{equation}
The mean square displacement over an interval $\tau$ is then related to the correlation function by
\begin{equation}\label{expand}
\langle [X(t)-X(t+\tau)]^2\rangle = 2 \langle X^2 \rangle- 2\Xi(\tau)
\end{equation}
 
The Planckian random walk described above leads over short intervals to a mean square displacement linear in $\tau$:
\begin{equation}\label{shortinterval}
\langle [X(t)-X(t+\tau)]^2\rangle =  c^2 t_P \tau(2/\pi),
\end{equation}
where we have normalized the coefficient to agree with the value of $\Delta X^2= c\tau \ell_P' = c^2 \tau t_P (2/\pi)$ derived above from the wavepacket theory normalized to black hole entropy.
It is expected that the simple random-walk described by Eq. (\ref{shortinterval}) should hold for $\tau<< 2L/c$, since the size of the apparatus should not affect the behavior. 

 For $c\tau= 2L$,  the autocorrelation must vanish, because the random walk in phase is limited by the size of the apparatus.  The light in the two directions of the interferometer is not  the same as waves arriving from infinity, but is prepared differently, by  interactions with the beamsplitter.  The beamsplitter has a definite (classical) position at any given time; however, the light from this one instant enters the detector at times separated by $2L/c$, having propagated in different directions.
The random walk is thus bounded; an interferometer does not measure holographic fluctuations of larger physical size, but only those within the causal boundaries defined by a single light round trip    $\tau= 2L/c$, the longest time interval over which relative phases in the two directions experience a differential random walk that affects the measured phase. If one arm is regarded as  a reference clock, the train of pulses used to compare with the other arm only has a ``memory'' lasting for a time $2L/c$ before it is ``reset''.

These constraints lead to an estimate of the overall correlation function that is sufficiently precise to design an exploratory experiment.  The total variance is   $\langle X^2\rangle= \Xi(\tau=0) = 4 ct_PL/\pi$. Using Eqs.(\ref{expand}) and (\ref{shortinterval}), that is, simply extrapolating the linear behavior to  $\tau= 2L/c$,  the autocorrelation function then becomes
\begin{eqnarray}\label{timedomain} 
\Xi(\tau)&=  & (2ct_P/\pi) (2L - c\tau),    \qquad   0<c\tau<2L\\
&=  & 0,  \qquad  c\tau>2L.
\end{eqnarray}

The time-domain correlation fixes other measurable statistical properties, including the frequency spectrum.
The spectrum   $\tilde\Xi(f)$ is given by the cosine transform,
\begin{equation}
\tilde\Xi(f)= 2 \int_0^\infty d\tau \Xi(\tau) \cos(\tau\omega),
\end{equation}
where $\omega=2\pi f$.
Integration of this formula using Eq.(\ref{timedomain}) gives a prediction for the spectrum of the  displacement noise,
\begin{equation}\label{displacementspectrum}
\tilde\Xi(f)= {4c^2t_P\over \pi (2\pi f)^2}[1-\cos (f/f_c)], \qquad f_c\equiv c/4\pi L.
\end{equation}
The spectrum at  frequencies above $f_c$ oscillates with a decreasing envelope that scales like $\tilde\Xi(f)\propto f^{-2}$. At  frequencies much higher than $f_c$, the mean square fluctuation in a frequency band $\Delta f$ goes like $\tilde\Xi(f)\Delta f \propto (\Delta f/ f)(c^2 t_P/f)$.  This is independent of $L$, as it should be, and shows the increasing variance in position as $f $ decreases.

The apparatus size acts as a cutoff; fluctuations from longer longitudinal modes do not add to the fluctuations, and the spectrum at  frequencies far below $f_c$ approaches a constant.
In particular,  the mean square displacement averaged over a time $T$ much longer than $2L/c$ is
$ \approx (4ct_PL/\pi) (2L/cT)$,
showing what has already been stated, that the effect in a given spatial volume decreases in a time averaged experiment.
This simply reflects the fact that the frequency spectrum of the displacement is flat at frequencies far below the inverse system size.

These results can be extended to estimate the cross correlation for two interferometers, including the cases when they are slightly displaced from each other or misaligned.  Let $X_A,X_B$ denote the apparent arm length difference in each of two interferometers $A$ and $B$.
The cross correlation  is defined as the limiting average,
 \begin{equation}\label{xcorrelation}
\Xi(\tau)_\times\equiv \lim_{T\rightarrow\infty} (2T)^{-1}\int_{-T}^{T} dt X_A(t) X_B(t+\tau)\equiv\langle X_A(t)X_B(t+\tau)\rangle.
\end{equation}
 Based on the above interpretation of the uncertainty, we adopt the following rule for estimating cross correlations. Transverse  displacements are the same to first order on the spacelike surface defined by each null plane wavefront, and decorrelate only slowly (to second order in  $\omega$ for each mode) with transverse separation. Thus, the differential phase perturbations in the two machines are almost the same when both pairs of laser wavefronts are traveling in the same direction at the same time in the lab frame, with  small transverse separation compared to the propagation distance. If they are displaced or misaligned the correlation is reduced by  appropriate directional and overlap projection factors.  
 
 For example, consider  two aligned interferometers configured as in  Figure \ref{topview}(b), displaced by a small distance $\Delta L$ along one axis, where $\Delta L<<L$. The cross correlation of measured phase displacement (in length units) becomes
\begin{eqnarray}\label{align}
\Xi_{\times}(\tau)&\approx  & (2ct_P/  \pi) (2L-2\Delta L- c\tau),    \qquad   0<c\tau<2L-2\Delta L\\
&=  & 0,  \qquad  c\tau>2L-2\Delta L.
\end{eqnarray}
That is, the cross correlation is the same as the autocorrelation of the largest interferometer that would fit into the in-common spacetime volume between the two. These formulae provide concrete predictions for experimental tests of the framework presented here. Assuming the theory is correctly normalized by black hole thermodynamics, there are no free parameters in the predictions, so there is a clearly defined experimental target. 

Another simple configuration is two adjacent interferometers, with one arm of each parallel and adjacent to the other but with the other arms extending in opposite directions, as in Figure \ref{topview}(c). In this setup the spacelike surfaces defined by wavefronts in the opposite arms  never coincide. Since the causal diamonds of those end mirrors do not overlap, the holographic noise in the two signals is uncorrelated, even though their other diamonds do overlap.

In addition, in this configuration the beamsplitters are at right angles to each other and therefore the phases of reflected light depend on precisely orthogonal components of displacement, so their signals should be uncorrelated. 
This result can be derived in the operator description. For the configuration just described, with opposite arms along axis 1, the cross correlation of the two machines $A$ and $B$ at zero lag ($\tau= 0$)  is
\begin{eqnarray}\label{oppositearm}
\langle X_A X_B\rangle&=  & \langle [-x_{1A}(t)-x_{2A}(t-2L/c) ][x_{1B}(t)-x_{2B}(t-2L/c)]\rangle\\
&=  & \langle - x_{1A}(t)x_{1B}(t)+x_{2A}(t-2L/c)x_{2B}(t-2L/c)\label{cancelterms}\\  
&&-x_{2A}(t-2L/c)x_{1B}(t)+x_{1A}(t)x_{2B}(t-2L/c))\rangle.\label{zeroterms}
\end{eqnarray}
In machine $A$, a positive displacement along axis 1 lengthens arm 1, while in machine $B$ it shortens it; this appears as the opposite signs for the machines in line (\ref{oppositearm}).
The terms in line (\ref{cancelterms}) then cancel, while the summed terms in line (\ref{zeroterms}) average to zero by symmetry, so the overall  cross correlation vanishes.
This argument also shows that the cross correlation in this setup should vanish, providing a useful configuration for an experimental null control. Note that cross correlation in this setup would not vanish for fluctuations caused by gravitational waves or metric fluctuations, or  other sources of conventional environmental noise.

\section{comparison with experiments}

It is useful to compare sensitivity to Planckian directional position fluctuations in the language of frequency error (or Allan variance) often used to characterize clocks.  Planck precision here does not mean a clock error of one Planck time; rather, it means a total random error  that accumulates like a random walk of a Planck time per Planck time. Variance  from an ideal clock (or between two clocks) grows linearly with time interval $\tau$, while the fractional clock error decreases over longer intervals like $\tau^{-1/2}$.  In the case of Planckian position noise, the difference of position in two directions similarly fluctuates as a Planckian random walk up to the scale of the apparatus.

With the adopted Planckian normalization (Eq. \ref{shortinterval}), the fractional standard deviation over a duration $\tau$ is
\begin{equation}\label{planckian}
{\Delta \nu (\tau)\over \nu} \approx \Delta t (\tau)/\tau = \sqrt{ 2\times 5.39\times 10^{-44}{\rm sec}\over \pi \tau} = 1.85\times 10^{-22}/\sqrt{\tau/\rm{sec}}.
\end{equation}
For comparison,  
 frequency error  in the best atomic clocks is currently \cite{chou} $
{\Delta \nu (\tau)/ \nu}=2.8\times 10^{-15}/\sqrt{\tau/\rm{sec}}$.
Thus the predicted  noise level is far below the currently practicable level of time measurements using atomic clocks. It is not possible for example to measure Planckian position variations using  local time standards.

\begin{figure}[b]
 \epsfysize=5in 
\epsfbox{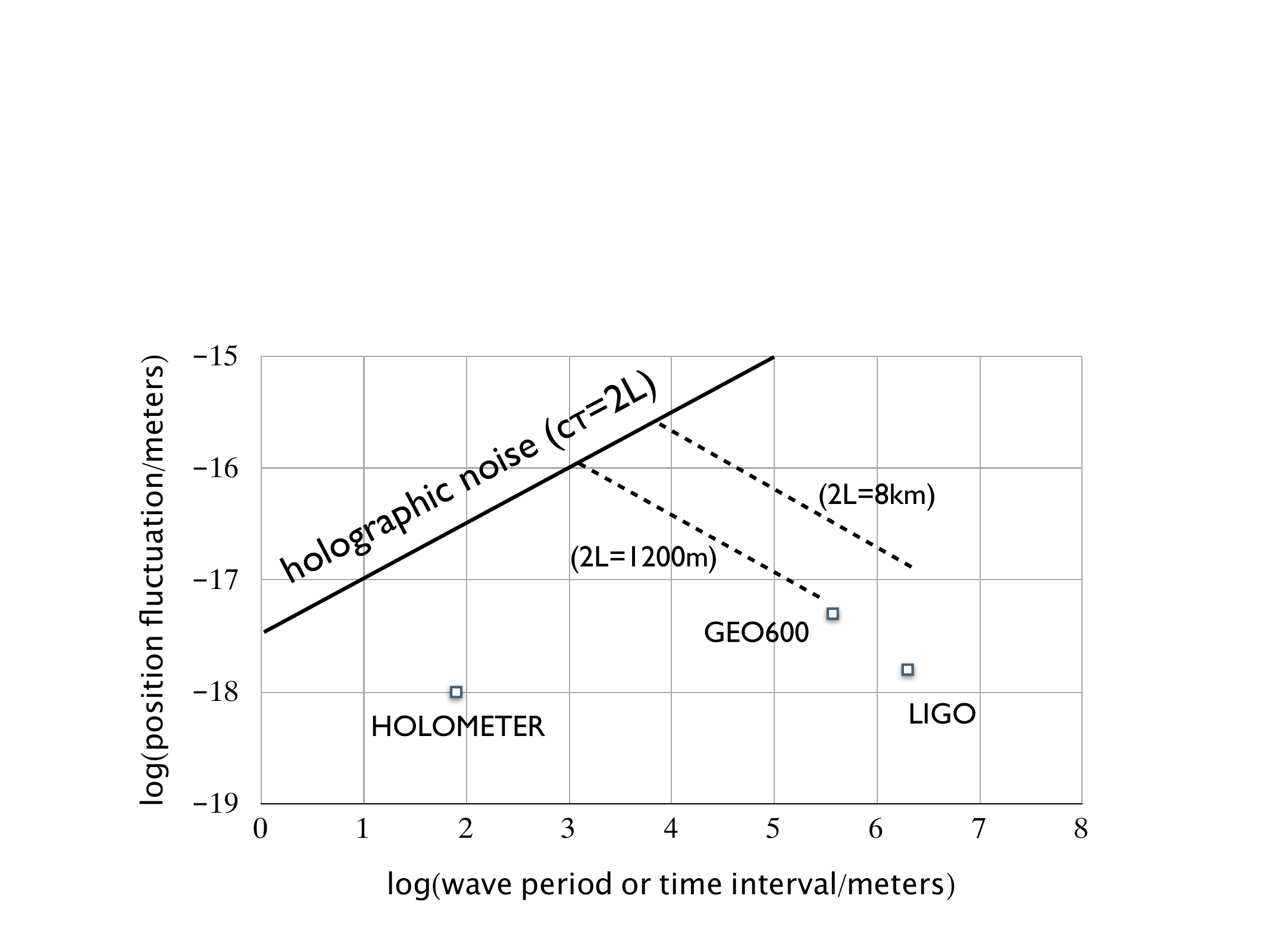} 
\caption{ \label{GEOLIGO}
Differential position fluctuation as a function of time interval or wave period.  Both scales are in meters.  For interferometers, the radius of the causal diamond is the arm length, $L$.     The holographic noise line refers to the transverse displacement amplitude estimated in Eq.(\ref{planckian}), for  time interval $c\tau=2L$.  For averaging time $c\tau>>2L$, corresponding to the flat low-frequency limit of the predicted spectrum, Eq. (\ref{lowfrequency}), the mean fluctuation amplitude falls off as $\sqrt{\tilde\Xi(f)\Delta f}\approx \sqrt{2L/c\tau}$, as shown by the dashed lines for the 600 meter arms of GEO600 and 4km arms of LIGO. 
Current    GEO600 and LIGO sensitivities show  the  standard deviation of   displacement  at the minimum of their noise curves, about   800 Hz and 150 Hz  respectively, for gravitational wave induced displacements.   This plot does not show the additional factor to correct for the reduced response of these particular layouts to holographic noise.  Folded arms in GEO600, and Fabry-Perot arm cavities in LIGO,   reduce  sensitivities to holographic noise by about a factor of 2 and 100 respectively at low frequencies, so current measurements remain  above the holographic noise predictions.
The point labeled Holometer shows the estimated photon-shot-noise limit for two 40-meter, co-located interferometers, with the same cavity power as GEO600, cross-correlated for 1 hour up to frequency $ c/2L= 3.74\  {\rm MHz}$. An instrument of this kind should be able to convincingly rule out or detect holographic noise.}
\end{figure}

\begin{figure}[b]
 \epsfysize=5in 
\epsfbox{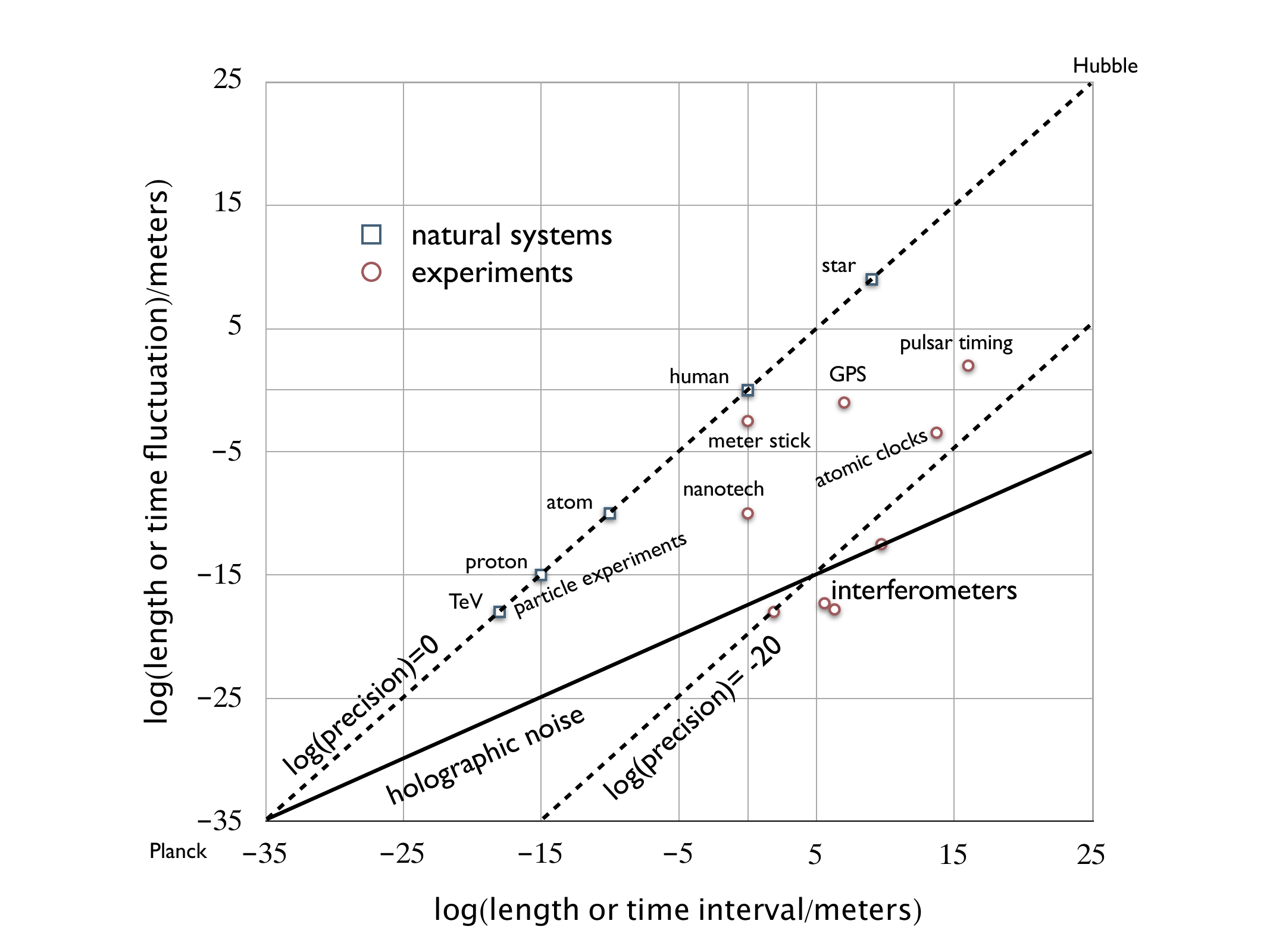} 
\caption{ \label{experimentsall}
Comparison of measurement precision of a larger variety of experiments to position fluctuations. Differential position change in length or time is plotted as a function of system size or duration, both with decimal log scales in meters, extending from the Planck scale to the Hubble scale.    The holographic noise prediction and interferometer sensitivities are shown as in Figure (\ref{GEOLIGO}), with the addition of LISA. Rough estimates of precision with current technology for other experimental techniques are labeled. The upper dashed line shows rough scales of natural systems;  the lower dashed line shows a fractional transverse position fluctuation, or angular indeterminacy, $\Delta x/x= \Delta\tau/\tau=\Delta\theta= 10^{-20}$. Laser-based  interferometry is the most sensitive technique by this measure, and the only one currently capable of detecting holographic noise.}
\end{figure}

 However, over short (but still macroscopic) time intervals, Planckian  noise in position differences, between two directions, may be  detectable using
interferometers. For durations $\tau \approx 2L/c$, interferometers are, in this limited differential sense, by far the most stable clocks.  
Current sensitivities of LIGO and GEO600 are shown in Figure (\ref{GEOLIGO}), along with the holographic noise prediction, Eq. (\ref{planckian}).   Figure (\ref{experimentsall}) compares   a wider range of experimental approaches, and shows that interferometry is currently the most promising approach to detect the effect. 
 
Consider first a direct, na\"ive comparison of  predicted and measured displacement spectral densities. The low-frequency limit of the predicted  spectral density,  Equation(\ref{displacementspectrum}), is
 \begin{equation}\label{lowfrequency}
\tilde\Xi(f)\approx  {8t_PL^2\over \pi }, \qquad f<<f_c.
\end{equation}
The rms fluctuation corresponding to this flat spectrum is shown in Figure (\ref{GEOLIGO}) as dashed lines for LIGO and GEO600.  The  experimental points are taken from published noise curves\cite{Abbott:2009ws,Luck:2010rt,GEO600noise} at the most sensitive frequency.  For $L=600$m, the predicted amplitude spectral density is $\sqrt{\tilde\Xi(f)}= 2.2\times 10^{-19}\ {\rm m}/\sqrt{\rm Hz}$, slightly higher than the observed minimum noise in GEO600. For LIGO, the predicted value is much higher than the measured noise.
 
 At first glance this comparison makes it look like holographic noise should already have been detected, if it exists.
The fact that LIGO does not see excess noise at this level constrains the spectral density of random noise in metric fluctuations to well below the Planckian value.    Indeed, this result  rules out some earlier theories of Planck-scale fluctuations\cite{AmelinoCamelia:2001dy}.  
 
  However, it is important  to take detailed account of the response of these specific interferometers to holographic displacements at the frequencies being measured.   The quoted  noise levels are for  displacements caused by gravitational waves, which have a different physical character from holographic noise. LIGO and GEO600 both employ interferometer configurations that increase their sensitivities to low frequency gravitational waves, without increasing their sensitivity to holographic noise.
 
  In the case of GEO600, the arms are folded; in the case of LIGO, the arms have Fabry-Perot cavities. In both cases, extra inboard mirrors near the beamsplitter reflect light back to the end mirrors.  These  features   amplify the phase response to low frequency gravitational-wave displacements.  Total phase displacement from gravitational waves adds coherently  over multiple reflections in the arm folds or cavities as the wave passes, just as if the arms were longer. The displacement of the inboard mirrors by a low frequency wave adds to the measured phase displacement. However, such single-arm amplification  does not happen in response to holographic position jitter,  since the jitter does not affect  normal reflections along a single direction, but only arises in a comparison of two different directions.  The inboard test masses (mirrors) and end mirrors reflect light in a single direction, and in each arm they are always in the  position eigenstate for that direction. Indeed, this behavior derives directly from the definition of position we adopted at the beginning: {\em everything} along a null trajectory in a single direction is always in the same position state for that direction, and all interactions with light traveling in that direction are  the same as the classical case. If the transverse jitter is visualized as a classical motion, this directional coherence appears like a quantum-mechanical ``spooky action at a distance''.
But the only departure from classical behavior comes where the positions in two directions are compared, in this case by beamsplitter reflections. 
    
  Thus, the extra reflections in cavities or folded arms do not contribute holographic phase noise. 
  In these configurations, the  signal  is  sensitive  to holographic jitter only of the beamsplitter relative to the end mirrors at any given time. That displacement is determined by the physical length of the arms, i.e., the causal diamond,  which is smaller than the light storage time that determines the the LIGO and GEO600 low-frequency gravitational-wave sensitivity.   Thus, LIGO's holographic-noise sensitivity is worse than its sensitivity to gravitational wave displacements at low frequencies by about the number of reflections in the Fabry-Perot cavities,  or about a factor of a hundred; in the case of GEO600, the suppression from folded arms is a factor of 2. When this factor is included, holographic noise is not currently ruled out by either LIGO or GEO600.  The latter is within a factor of two of being limited primarily by holographic noise, if the overall normalization adopted here is correct.

It appears that current interferometer technology is nearly able to detect the effect, but that a new experiment must be built to achieve a convincing detection or limit.    The optimal  frequency for holographic noise detection is  $\approx c/2L$, two to three orders of magnitude higher than the optimal frequencies of gravitational wave detectors. The   design should be optimized  to allow a direct measurement of holographic noise, and to distinguish it from other noise sources, particularly the dominant photon shot noise. 

\begin{figure}[b]
 \epsfysize=5in 
\epsfbox{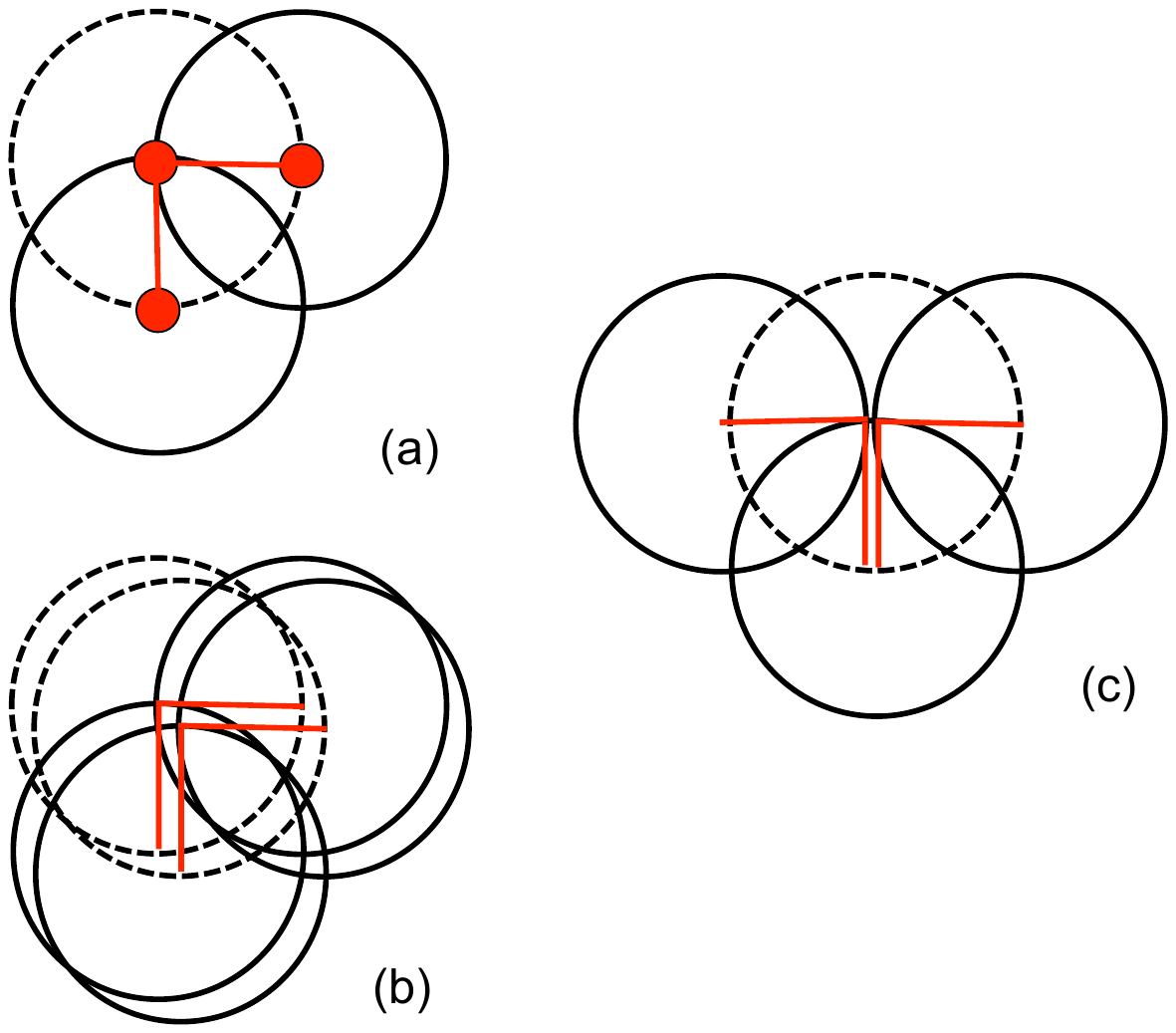} 
\caption{ \label{topview}
(a) Projection of the causal diamond in Figure (\ref{interferometerdiamond}) onto the plane of an interferometer. The time axis is not shown. The causal spheres around the two end mirrors are also shown, as solid circles. Beamsplitter position (center of dashed circle)  ``jitters'' in the directions transverse to laser wavefronts coming from the two end mirrors, along the tangent directions of each of their causal spheres. 
Holographic noise appears in the dark-port signal that measures the position difference of the beamsplitter in the two directions. 
   (b) Projection showing two interferometers slightly displaced from each other. Most of the spacetime volume overlaps. The Hilbert spaces of the their diamonds are highly entangled, leading to highly correlated holographic noise signals, as in Eq. (\ref{align}).
(c) Projection showing two interferometers with one nearly-collocated arm, with other arms extending in opposite directions. The causal diamonds from the left and right end mirrors do not overlap at all, so the holographic-noise part of their signals  display zero correlation. This null configuration is a useful control  for experiments.}
\end{figure}

  One way to isolate the holographic component of  noise as a distinctive signal is to cross-correlate  two nearly-co-located interferometers at high frequencies.  Because of their overlapping spacetime volumes, their holographic displacements are correlated (as in Figure \ref{topview}(b) and Eq. \ref{align}), whereas their photon shot noise is independent.  With  a long integration, a time-averaged holographic correlation  emerges above uncorrelated photon shot noise, in a way similar to the correlation technique used with LIGO at much lower frequencies for isolating gravitational-wave stochastic backgrounds. (The LIGO correlation studies however do not themselves constrain holographic noise, because the interferometers being correlated are not co-located--- indeed, they are kept separate to avoid acoustic sources of cross correlation at low frequency.)

 An experiment based on this concept is currently under construction at Fermilab using two nearly co-located interferometers with 40-meter arms. Their signals will be correlated at high frequencies, that is, $\approx c/2L\approx 3.74\  {\rm MHz} (40{\rm m}/ L)$, to reduce shot noise and distinguish other external sources of cross correlation.   
 If  noise is dominated by photon shot noise  comparable to GEO600 (that is, if they have the same laser power on the beamsplitters), the sensitivity can be estimated by extrapolation from GEO600's measured noise at $\approx 800$ Hz. The differential-position amplitude spectral density in $\rm{m/\sqrt{ Hz}}$ is  the same; the rms displacement sensitivity is worse than GEO600 by the bandwidth factor $\sqrt{3.74\times10^6/800}$, but is then improved over an integration interval $\tau$ by a factor $(\tau \times 3.74\times10^6 {\rm Hz})^{-1/4}$. This  estimate  for $\tau=$ 1 hour is labeled ``Holometer'' in Figure \ref{GEOLIGO}.  An experiment based on  this design that achieves the photon shot noise limit  should achieve   a highly significant detection of Planckian holographic noise, if it exists. As a control, the holographic noise can be ``turned off''  by correlating interferometers in a null configuration,  as in Figure \ref{topview}(c).

  This experiment will explore quantum departures from classical behavior of position  in spacetime that have never been tested before to Planckian precision, and that lie beyond the current predictive scope of reliably tested physical theory. Because new spacetime physics is suspected to appear at the Planck scale, it appears to be well motivated as an exploratory experiment.

\acknowledgments
I am grateful to  D. Berman, A. Chou, and M. Perry  for useful  comments and discussions, and to the Aspen Center for Physics for hospitality. This work was supported by the Department of Energy at Fermilab under Contract No. DE-AC02-07CH11359, and by NASA grant NNX09AR38G at the University of Chicago.

\break


\begin{thebibliography}{}



\bibitem{Banks:2011av}
  T.~Banks,
  ``Holographic Space-Time: The Takeaway,''
  arXiv:1109.2435 [hep-th].

\bibitem{Banks:2011jt}
 T.~Banks and J.~Kehayias,
  ``Fuzzy Geometry via the Spinor Bundle, with Applications to Holographic
  Space-time and Matrix Theory,''
  arXiv:1106.1179 [hep-th].
  
 \bibitem{wigner}
E. P. Wigner, ``Relativistic Invariance and Quantum Phenomena'', Rev. Mod. Phys. {\bf 29}, 255 (1957)

\bibitem{salecker}
H. Salecker \& E. P. Wigner, ``Quantum Limitations of the Measurement of Space-Time Distances'', Phys. Rev. {\bf 109}, 571 (1958)


\bibitem{peres}
A. Peres, ``Measurement of Time by Quantum Clocks'', Am. J. Phys. {\bf 48}(7), 552 (1980)

\bibitem{braginsky} V. B. Braginsky and F. Ya. Khalili, ``Quantum Measurement'',  Cambridge:  University Press (1992)


\bibitem{aharonov}
Y. Aharonov, J. Oppenheim, S. Popescu, B. Reznik, W. G. Unruh, ``Measurement of time of arrival in quantum mechanics'', Phys. Rev. A {\bf 57}, 4130  (1998)

\bibitem{zurek} W. H. Zurek, ``Decoherence,  einselection, and the quantum origin of the classical'', Rev. Mod. Phys. {\bf  75}, 715 (2003)

\bibitem{Padmanabhan:1987au}
  T.~Padmanabhan,
  ``Limitations on the operational definition of space-time events and quantum gravity'',
  Class.\ Quant.\ Grav.\  {\bf 4}, L107 (1987)

\bibitem{'tHooft:1993gx}
  G.~'t Hooft,
  ``Dimensional reduction in quantum gravity,''
  in ``Conference on Particle and Condensed Matter Physics (Salamfest)'',
  edited by A. Ali, J. Ellis, and S. Randjbar-Daemi (World Scientific, Singapore, 1993),
  [arXiv:gr-qc/9310026]

\bibitem{Susskind:1994vu}
  L.~Susskind,
  ``The World As A Hologram,''
  J.\ Math.\ Phys.\  {\bf 36}, 6377 (1995)
  
\bibitem{Jacobson:1995ab}
  T.~Jacobson,
  ``Thermodynamics of space-time: The Einstein equation of state,''
  Phys.\ Rev.\ Lett.\  {\bf 75}, 1260 (1995)

\bibitem{Bousso:2002ju}
 R.~Bousso,
 ``The holographic principle,''
 Rev.\ Mod.\ Phys.\  {\bf 74}, 825 (2002)
 

  
\bibitem{Padmanabhan:2009vy}
  T.~Padmanabhan,
  ``Thermodynamical Aspects of Gravity: New insights,''
  Rept.\ Prog.\ Phys.\  {\bf 73}, 046901 (2010)
  
\bibitem{Verlinde:2010hp}
  E.~P.~Verlinde,
  ``On the Origin of Gravity and the Laws of Newton,''
  JHEP {\bf 1104}, 029 (2011)
  [arXiv:1001.0785 [hep-th]].
  
\bibitem{Seiberg:1999vs}
  N.~Seiberg and E.~Witten,
  ``String theory and noncommutative geometry,''
  JHEP {\bf 9909}, 032 (1999)
  
\bibitem{connes}
A. Connes, M. Marcolli, ``A Walk in the Noncommutative Garden'', 
[arXiv:math/0601054] 

\bibitem{connesbook}
A. Connes, ``Noncommutative Geometry'',  Academic Press (1994)


\bibitem{mattingly}
D. Mattingly, 
``Modern Tests of Lorentz Invariance", 
Living Rev. Relativity 8,  (2005),  5
(http://relativity.livingreviews.org/Articles/lrr-2005-5/)

    

\bibitem{AmelinoCamelia:2009zz}
  G.~Amelino-Camelia, C.~Laemmerzahl, F.~Mercati and G.~M.~Tino,
  ``Constraining the Energy-Momentum Dispersion Relation with Planck-Scale
  Sensitivity Using Cold Atoms,''
  Phys.\ Rev.\ Lett.\  {\bf 103}, 171302 (2009)
  
  
 \bibitem{fermi2009}
  A. A. Abdo et al., ``A limit on the variation of the speed of light arising from quantum gravity effects'', Nature {\bf 462}, 331-334,  doi:10.1038/nature08574 (2009)
  
\bibitem{Laurent:2011he}
  P.~Laurent, D.~Gotz, P.~Binetruy, S.~Covino and A.~Fernandez-Soto,
  ``Constraints on Lorentz Invariance Violation using INTEGRAL/IBIS
  observations of GRB041219A,''
  arXiv:1106.1068 [astro-ph.HE].
  
\bibitem{Ellis:1983jz}
  J.~R.~Ellis, J.~S.~Hagelin, D.~V.~Nanopoulos and M.~Srednicki,
  ``Search For Violations Of Quantum Mechanics,''
  Nucl.\ Phys.\  B {\bf 241}, 381 (1984)
  
\bibitem{AmelinoCamelia:1999gg}
  G.~Amelino-Camelia,
  ``Gravity-wave interferometers as probes of a low-energy effective  quantum
  gravity,''
  Phys.\ Rev.\  D {\bf 62}, 024015 (2000)
  
\bibitem{AmelinoCamelia:2001dy}
  G.~Amelino-Camelia,
  ``A phenomenological description of quantum-gravity-induced space-time
  noise,''
  Nature {\bf 410}, 1065 (2001)

     
\bibitem{Schiller:2004tf}
  S.~Schiller, C.~Laemmerzahl, H.~Mueller, C.~Braxmaier, S.~Herrmann and A.~Peters,
   ``Experimental limits for low-frequency space-time fluctuations from
  ultrastable optical resonators,''
  Phys.\ Rev.\  D {\bf 69}, 027504 (2004)

\bibitem{Smolin:2006pa}
  L.~Smolin,
 ``Generic predictions of quantum theories of gravity,''
  [arXiv:hep-th/0605052]
  

\bibitem{Ng:2000fq}
  Y.~J.~Ng,
  ``From computation to black holes and space-time foam,''
  Phys.\ Rev.\ Lett.\  {\bf 86}, 2946 (2001)
  [Erratum-ibid.\  {\bf 88}, 139902 (2002)]
  [arXiv:gr-qc/0006105]
  
\bibitem{Ng:2004xr}
  Y.~J.~Ng,
  ``Quantum foam and quantum gravity phenomenology,''
  Lect.\ Notes Phys.\  {\bf 669}, 321 (2005)
  


\bibitem{Lloyd:2005dm}
  S.~Lloyd,
  ``Quantum limits to the measurement of spacetime geometry,''
  [arXiv:quant-ph/0505064]

\bibitem{caves1980}
C. ~ Caves, ``Quantum-Mechanical Radiation-Pressure Fluctuations in an Interferometer'', Phys. Rev. Lett. {\bf 45}, 75 (1980)

 
\bibitem{Hogan:2007pk}
  C.~J.~Hogan,
  ``Measurement of Quantum Fluctuations in Geometry''
  Phys. Rev. D 77, 104031 (2008)
  
\bibitem{Hogan:2008zw}
  C.~J.~Hogan,
  ``Indeterminacy of Quantum Geometry''
  Phys Rev D.78.087501 (2008)

\bibitem{Hogan:2008ir}
  C.~J.~Hogan and M.~G.~Jackson,
  ``Holographic Geometry and Noise in Matrix Theory,'' Phys. Rev. D.79.124009 (2009)

  
\bibitem{Hogan:2009mm}
  C.~J.~Hogan,
  ``Holographic Noise in Interferometers,''
  [arXiv:0905.4803]

      
\bibitem{chou}
 C.-W. Chou, D.B. Hume, J.C.J. Koelemeij, D.J. Wineland, and T. Rosenband, ``Frequency Comparison of Two High-Accuracy Al+ Optical Clocks'', Physical Review Letters, 104, 070802 (2010)




\bibitem{Abbott:2009ws}
  B.~P.~Abbott {\it et al.}  [LIGO Scientific Collaboration and VIRGO
                  Collaboration],
  ``An Upper Limit on the Stochastic Gravitational-Wave Background of
  Cosmological Origin,''
  Nature {\bf 460}, 990 (2009)


\bibitem{Luck:2010rt}
  H.~L\"uck {\it et al.},
  ``The upgrade of GEO600,''
  J.\ Phys.\ Conf.\ Ser.\  {\bf 228}, 012012 (2010)
  [arXiv:1004.0339]
  
  \bibitem{GEO600noise}
LIGO Scientific Collaboration, ``A gravitational wave observatory operating beyond the quantum shot-noise limit'', Nature Physics, 11 September 2011 (http://dx.doi.org/10.1038/nphys2083)
  
\end{thebibliography}
\end{document}